\documentclass[aps,pra,twocolumn,groupedaddress,floatfix,superscriptaddress,numbers]{revtex4-2}

\usepackage{lipsum}  
\usepackage[english]{babel}
\usepackage{bbold}
\usepackage{enumitem}
\usepackage[latin9]{inputenc}
\usepackage{braket}
\usepackage{amsfonts}
\usepackage{amsmath,amssymb,amsthm,mathrsfs,amsfonts,dsfont}
\usepackage{mathtools}
\usepackage{cancel}
\usepackage{color}
\usepackage{graphicx}
\usepackage{comment}
\usepackage{multirow}
\usepackage[colorlinks = true]{hyperref}
\usepackage{bbm}
\usepackage{cool}

\newcommand{\mr}[1]{\mathrm{ #1}}

\newcommand{\mirroredpropto}{\mathrel{\reflectbox{\rotatebox[origin=c]{0}{$\propto$}}}}
\newcommand{\Tgate}{T}

\usepackage{varioref}
\labelformat{equation}{#1}

\labelformat{section}{#1}

\labelformat{figure}{#1}

\labelformat{proposition}{#1}

\labelformat{lemma}{#1}

\labelformat{theorem}{#1}

\labelformat{observation}{#1}

\labelformat{definition}{#1}

\emergencystretch=\maxdimen
\usepackage{ytableau}
\ytableausetup{boxsize = 3pt}

\newsavebox{\mstrut}
\newcommand{\bbra}[1]{%
    \sbox{\mstrut}{\(#1\)}%
    \mathinner{\langle\kern-0.3\ht\mstrut\left\langle{#1}\right|}%
}
\newcommand{\kett}[1]{%
    \sbox{\mstrut}{\(#1\)}%
    \mathinner{\left|{#1}\right\rangle\kern-0.3\ht\mstrut\rangle}%
}

\newcommand{\thx}{\thanks{These authors contributed equally.}}
\newcommand{\itp}{\affiliation{Institute for Theoretical Physics, University of Innsbruck, 6020 Innsbruck, Austria}}
\newcommand{\iqoqi}{\affiliation{Institute for Quantum Optics and Quantum Information of the Austrian Academy of Sciences, 6020 Innsbruck, Austria}}
\newcommand{\planqc}{\affiliation{PlanQC GmbH, 85748 Garching, Germany}}
\newcommand{\chicago}{\affiliation{Pritzker School of Molecular Engineering, University of Chicago, Chicago, Illinois 60637, USA}}

\begin{document}
\title{Dark spin-cats as biased qubits}
\date{\today}

\author{Andreas Kruckenhauser} \thx\itp\iqoqi\planqc 
\author{Ming Yuan} \thx\chicago 
\author{Han Zheng} \thx\chicago
\author{Mikhail Mamaev} \chicago
\author{Pei Zeng} \chicago
\author{Xuanhui Mao} \chicago
\author{Qian Xu} \chicago
\author{Torsten V. Zache} \itp \iqoqi
\author{Liang Jiang} \chicago
\author{Rick van Bijnen} \itp \iqoqi \planqc
\author{Peter Zoller} \itp \iqoqi

\begin{abstract}We present a biased atomic qubit, universally implementable across all atomic platforms, encoded as a `spin-cat' within ground state Zeeman levels. The key characteristic of our configuration is the coupling of the ground state spin manifold of size $F_g \gg 1$ to an excited Zeeman spin manifold of size $F_e = F_g - 1$ using light. This coupling results in eigenstates of the driven atom that include exactly two dark states in the ground state manifold, which are decoupled from light and immune to spontaneous emission from the excited states. These dark states constitute the `spin-cat', leading to the designation `dark spin-cat'. We demonstrate that under strong Rabi drive and for large $F_g$, the `dark spin-cat' is autonomously stabilized against common noise sources and encodes a qubit with significantly biased noise. Specifically, the bit-flip error rate decreases exponentially with $F_g$ relative to the dephasing rate. We provide an analysis of dark spin-cats, their robustness to noise, and discuss bias-preserving single qubit and entangling gates, exemplified on a Rydberg tweezer platform.
\end{abstract}

\maketitle

Development of quantum computing hardware faces the requirements of scalability to large qubit numbers, while maintaining high levels of control and low error rates. 
To protect qubits from errors generated by imprecise control and environmental interactions, fault-tolerant quantum computing employs redundant encoding in logical qubits built from many physical qubits to detect and correct errors~\cite{Shor1996Fault,steane1999efficient,fowler2012surface,gottesman2013fault,campbell2017roads,gottesman2022opportunities}. 
However, the large resource cost of fault-tolerant quantum computing poses a significant challenge for present quantum hardware.
To mitigate this cost, one approach is to find different hardware-level encodings to suppress physical errors to higher order~\cite{gottesman2001encoding,mirrahimi2014dynamically,michael2016new}, introducing error-bias~\cite{aliferis2008fault,guillaud2019repetition,puri2020bias,chamberland2022building,albert_pair-cat_2019,yuan_construction_2022} or detect leakage errors~\cite{teoh2023dual,wu2022erasure, kubica2023erasure}. Encodings with strong error bias are particularly alluring, as they enable efficient quantum error correction (QEC) schemes with high permissible threshold for logical errors~\cite{tuckett2018ultrahigh, puri2020bias}.

\begin{figure}[!t] 
\center
    \includegraphics[width=1\columnwidth]{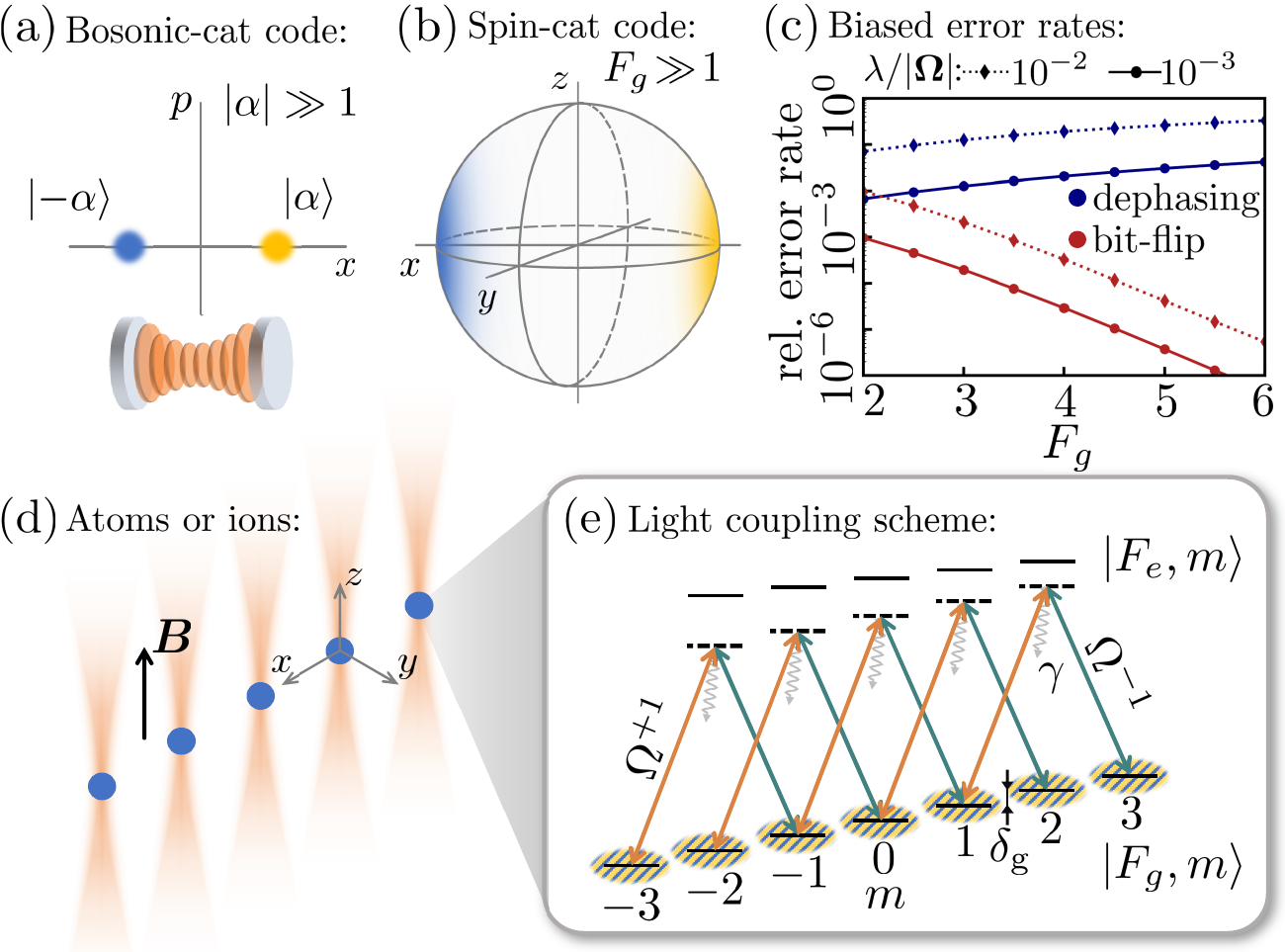}
\caption{Cat code:
(a) Phase space illustration of bosonic coherent states \mbox{$|\!\pm\!\alpha\rangle$}, with displacement $|\alpha|\!\gg\! 1$, realising a cat code, with \emph{e.g.} photons in a cavity.  
(b) Wigner distribution of spin coherent states realizing a biased spin-cat code for $F_g\!\gg\!1$. 
(c) Error rates due to colored noise with correlation time $1/\lambda$, relative to noise strength, are highly biased (bit-flip error decreases exponentially with $F_g$, while dephasing error only increases polynomially), and suppressed by increasing the laser Rabi frequency $\Omega$.
(d) Spin-cat codes are realizable with atomic platforms, (e) by coupling Zeeman-split spin manifolds with circularly polarized light $\Omega^q$ ($q\!=\!\pm 1$).
The spin coherent states of (b) are DSs in the $F_g$-manifold and distributed balanced over the magnetic sublevels $m$ (mixed blue/yellow coloring), giving rise to a biased error model.
}
\label{fig:1}
\end{figure}

In light of remarkable advances with atomic quantum computing, including scaling to a large qubit count \cite{scholl2021quantum, park2021cavity, ebadi2021quantum, kiesenhofer2023controlling, manetsch2024tweezer}, high fidelity gates \cite{gaebler2016high, finkelstein2402universal, anand2024dual, peper2024spectroscopy, wright2019benchmarking, gaebler2016high, ballance2016high, clark2021high, ma2023high, levine2019parallel, sorensen1999quantum, tsai2024}, and early fault-tolerant quantum computation \cite{pogorelov2024experimental, bluvstein2024logical}, we describe below a robust and biased qubit implementable across all atomic platforms. 
In analogy to the bosonic-cat code, we consider encoding a qubit as a `spin-cat'  in a Zeeman $F_{g}$-manifold $\ket{F_{g},\,m}$, with $m\! =\!-F_{g},\ldots,+F_{g}$, e.g. in a long-lived \mbox{(hyperfine-)} spin manifold of states (see Fig.\,\ref{fig:1}). While Ref.~\cite{omanakuttan2024fault} discussed `spin-cat' encoding for `bare' Zeeman levels, the defining feature of the present setup is that we couple the ground state manifold $F_{g}$ to an excited manifold $F_{e}\!=\!F_{g}\!-\!1$ with light, so that the eigenstates of the driven atom contain exactly two dark states (DSs) in the $F_g$-manifold decoupled from light and unaffected by spontaneous emission from the excited states. We identify these DSs with the spin-cat, hence the name `dark spin-cat'.  In contrast to Refs.~\cite{timoney2011quantum, aharon2013general, mikelsons2015universal, weidt2016trapped, aharon2017enhanced}, we are interested in the limit of large $F_{g}$.
For strong Rabi drive and large $F_g$, the `dark spin-cat' will be shown to be robust and autonomously stabilized against typical noise sources, and encode a qubit with highly biased noise. In particular, the bit-flip error rate is suppressed exponentially in $F_g$ as compared to the dephasing rate; Fig.\,\ref{fig:1}(c) shows a characteristic example of qubit robustness to `colored' noise. 
Below we discuss the unique noise resilience of "dark spin-cats", bias-preserving single qubit gates, and an illustration of entangling gates for the Rydberg (Ry) tweezer platform.

\textit{System.-} The atomic system we have in mind is illustrated in Fig.\,\ref{fig:1}(d-e). 
We consider two hyperfine (HF) manifolds $\ket{F_g,m}$, $\ket{F_e,m}$ with an overall energy splitting $\omega_{eg}$, dipole-coupled by laser or microwave radiation\,\footnote{Microwave radiation acts globally on the atom-ensemble, thus, lacks single site addressability. Single-site resolution can be restored using additional laser-driving fields. For simplicity we focus here on laser light.}.
A static magnetic field $\mathbf{B}$ (defining the quantization axis) generates a Zeeman splitting $\delta_{g,e}\!=\!g_{g,e}\mu_B|\mathbf{B}|$, where $g_{g,e}$ denote the manifolds' Land\'e-g factors and $\mu_B$ the Bohr magneton.
We consider atoms with a large (nuclear) spin ($F_{g,e}\!\gg\!1$) such as alkali or alkaline-earth atoms or ions, and light fields with polarization $q\in\{0,\,\pm1\}$, oscillation frequency $\omega_q$ and Rabi frequency $\Omega^q$ fulfilling the Raman resonance criteria $\omega_q\! -\! \omega_{q'} = \delta_g (q\!-\!q')$.

\begin{figure}[!t]
\center
    \includegraphics[width=1\columnwidth]{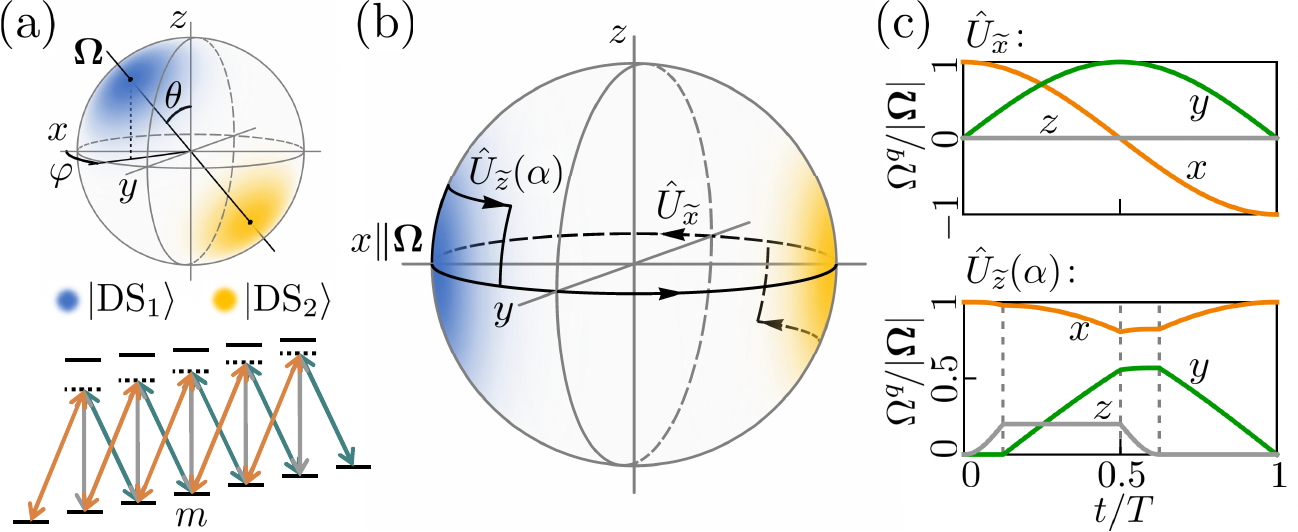}
\caption{(a) Wigner distribution of orthogonal `dark spin-cat' states on the Bloch sphere and associated light coupling scheme. 
(b) `Dark spin-cat' states on the equatorial plane and visualization of adiabatic single qubit gate trajectories. The associated parameter variations are displayed in panel (c), with $T$ denoting the gate time.}
\label{fig:2}
\end{figure}

In a rotating frame defined by the transformation \mbox{$\hat{\mathcal{U}} = \exp\{i[\omega_0 \hat{P}_e + \delta_g(\hat{F}_{e,z} + \hat{F}_{g,z})  ] t \}$}, the system Hamiltonian is
\begin{align}
\hat{H}_\mathrm{DS}/\hbar= -\Delta \hat{P}_e - \delta \hat{F}_{e,z}+\frac{1}{2}\sum_{q = 0,\pm 1}\!\! \left(\Omega^q \hat{\mathcal{C}}_q + \mathrm{h.c.}\right).
\label{eq:DSHamiltonian}
\end{align}
Here $\hat{F}_{g(e),i}$, with $i\!\in\!\{x,\,y,\, z\}$, are spin operators, $\hat{P}_e = \sum_m|F_e, m\rangle \langle F_e, m |$, while \mbox{$\Delta\!=\!\omega_0\!-\!\omega_{eg}$} denotes the overall detuning from the $F_e$-states, and \mbox{$\delta\! =\! \delta_g\! -\! \delta_e$} the differential Zeeman splitting. 
The light coupling is described by the operator \mbox{$\hat{\mathcal{C}}_q = \sum_m \mathcal{C}_{F_g,m; 1,q}^{F_e, m+q}|F_e,m\!+\!q\rangle \langle F_g, m |$}, with $\mathcal{C}_{F_g,m; 1,q}^{F_e,m+q}$ the Clebsch-Gordan (CG) coefficients incorporating dipole selection rules, see supplementary material (SM)\,\cite{SupMat}. 
Within the Wigner-Weisskopf approximation, spontaneous emission \cite{sobel2016introduction} from the $F_e$-states with rate $\gamma$ can be included as $-\Delta \rightarrow -\Delta -i\gamma/2$.

\textit{Dark states.-}
For the case of \mbox{$F_e \!=\! F_g\! -\!  1$} there are exactly two DSs, labelled $|\mathrm{DS}_{1,2}\rangle$, fulfilling 
\mbox{$\hat{H}_\mathrm{DS}|\mathrm{DS}_{1,2}\rangle = 0$} \cite{morris1983reduction}.
The two DSs are fully located in the $F_g$-manifold, thus immune to spontaneous emission, and are below identified as our qubit states.
The DSs are up to an orthonormalization given by two spin coherent states (SCSs)
$|\theta_{1,2},\,\varphi_{1,2}\rangle\!= \!\exp(-i\varphi_{1,2}\hat{F}_{g,z})\exp(-i\theta_{1,2}\hat{F}_{g,y})|F_g,\,F_g\rangle$.
The pairs of angles $(\theta_{1,2},\,\varphi_{1,2})$ are determined by the two relative phases and amplitudes of the Rabi frequencies $\Omega^q$. 
The two SCS are orthogonal, \emph{i.e.} on opposite sides of the Bloch sphere \mbox{$(\theta_2,\, \varphi_2) = (\pi\!-\!\theta_1,\,\pi\!+\!\varphi_1)$}, if the Rabi frequency's Cartesian components \mbox{$\Omega^x\! =\!(\Omega^{-1}\!-\! \Omega^{+1})/\sqrt{2}$}, \mbox{$\Omega^y\! =\! i(\Omega^{-1}\!+\! \Omega^{+1})/\sqrt{2}$}, $\mathrm{and}~\Omega^z = \Omega^0\notag$ are up to a global phase real-valued in the rotating frame defined by $\hat{\mathcal{U}}$. 
In this case, $\theta_1$ and $\varphi_1$ are given by the polar- and azimuthal angle of 
$\boldsymbol{\Omega} = \left(\Omega^x,\,\Omega^y,\,\Omega^z \right)$, respectively, see Fig.\,\ref{fig:2}(a) and SM\,\cite{SupMat}.

We now identify our `dark spin-cat' qubit with two SCSs pointing along the $\pm x$-axis
\begin{equation}\label{eq:encoding}
\begin{split}
&|\widetilde{0}\rangle \equiv e^{-i\frac{\pi}{2}\hat{F}_{g,y}}|F_g,\,F_g\rangle =  |\pi/2,0\rangle, 
~\mathrm{and}\\
&|\widetilde{1}\rangle \equiv e^{-i\frac{\pi}{2}\hat{F}_{g,y}}|F_g,\,-F_g\rangle =  e^{-i\pi F_g}|\pi/2,\pi\rangle,
\end{split}
\end{equation}
see Fig.\,\ref{fig:1}(b). 
The qubit states $|\widetilde{0}\rangle$ and $|\widetilde{1}\rangle$ are DSs when $\Omega^y,\Omega^z\! =\! 0$ and are identical to the maximally stretched spin-states along the $x$-axis.
The DSs subspace is unaffected by laser intensity and phase fluctuations when the drives originate from the same source.
States and operators labelled by $\widetilde{\cdot}$ are associated with the logical qubit states hereafter. 

A well-chosen adiabatic variation of laser parameters enables transporting the qubit states along trajectories on the Bloch sphere, with the two states always remaining antipodal. This allows for the implementation of a set of gates required for universal fault tolerant quantum computation~\cite{guillaud2019repetition}, where any errors will predominantly lead to dephasing (as discussed in detail below). Bit-flip errors are instead exponentially suppressed by the spin length $F_g$, if the states are antipodal.
In Fig.\,\ref{fig:2}(b-c) we show the time-dependent $\mathbf{\Omega}(t)$ sweeps and the corresponding $F_g$ Bloch sphere trajectories for a rotation around the $\widetilde{z}$-axis $\hat{U}_{\widetilde{z}}(\alpha)\!=\!\exp(-i\hat{\sigma}_{\widetilde{z}}\alpha/2)$ by an angle $\alpha$, and a $\pi$ rotation around the $\widetilde{x}$-axis $\hat{U}_{\widetilde{x}}\! =\! \exp(-i\hat{\sigma}_{\widetilde{x}}\pi/2)$. 
Here, $\hat{\sigma}_{\widetilde{x}}\! =\! \ket{\widetilde{1}}\bra{\widetilde{0}}\!+\!\mathrm{h.c.}$, $\hat{\sigma}_{\widetilde{y}}=-i(\ket{\widetilde{1}}\bra{\widetilde{0}}\!-\!\mathrm{h.c.})$, and
$\hat{\sigma}_{\widetilde{z}}\! =\! \ket{\widetilde{1}}\bra{\widetilde{1}}- \ket{\widetilde{0}} \bra{\widetilde{0}}$ are the `dark spin-cat' Pauli operators, thus $\widetilde{x},\,\widetilde{y}$, and $\widetilde{z}$ refer to the logical qubit Bloch sphere axes.
The implementation of $\hat{U}_{\widetilde{z}}(\alpha)$ is based on holonomic quantum processes \cite{zanardi1999holonomic, recati2002holonomic, albert2016holonomic, kam2023coherent}, with the enclosed area of the loop on the Bloch sphere determining the angle $\alpha$. 

\textit{Error-analysis.-}
Dominant external sources of noise for our system include fluctuating magnetic (electric) fields, laser intensity and phase fluctuations 
\footnote{
Additionally, imperfect laser polarization gives rise to AC-Stark shifts, altering the energy and the composition of the DSs. 
These effects are rendered negligible by the applied magnetic field \cite{aharon2013general}.
}, and non-magic trapping conditions for neutral atom quantum processors~\cite{le2013dynamical, thompson2013coherence}.
These sources of errors are described in the laboratory frame by low powers of spin operators $\prod_{i=x,y,z}(\hat{F}_{g,i})^{n_i}$ and typically $\sum_i n_i\!\leq\!2$~\cite{omanakuttan2024fault}. 
When transformed into the rotating frame defined by $\hat{\mathcal{U}}$, any off-diagonal elements of such noise operators acquire time-oscillating prefactors with frequency $\delta_g$, and can hence be suppressed by an external magnetic field $\mathbf{B}\!\mirroredpropto \!\delta_g$
\footnote{
Noise operators in the rotating frame: The diagonal components 
$\mathcal{\hat{U}}\big(\hat{F}_{g,z}\big)^n\mathcal{\hat{U}}^\dagger = \big(\hat{F}_{g,z}\big)^n$ are invariant under the rotating frame transformation.
The off-diagonal components \mbox{$\mathcal{\hat{U}}\,\hat{F}_{g,\pm}\,\mathcal{\hat{U}}^\dagger = \exp(\pm i\delta_g t)\hat{F}_{g,\pm}$} become oscillatory in time, with $\hat{F}_{g,\pm} \!=\! \hat{F}_{g,x}\! \pm\! i \hat{F}_{g,y}$.
Thus, off-diagonal noise components average out as long as the corresponding noise amplitude and time dependency is much smaller (slower) than $|\boldsymbol{\Omega}|$.
Note, for polynomials of spin operators, as considered in the main text, similar identities can be derived. 
}.
Together with the observation that diabatic effects during gate operation manifest also as $\hat{F}_{g,z}$ (in a suitably chosen frame, see SM\,\cite{SupMat}), this leaves longitudinal fields \mbox{$\sim\!(\hat{F}_{g,z})^{n_z}$} as the main remaining source of noise in the rotating frame \cite{aharon2013general}.
In the following we show that colored noise processes coupling to $\hat{F}_{g,z}$ with strength and spectral width smaller than $|\boldsymbol{\Omega}|$ lead to a reduced and biased error model in the limit $F_g\!\gg\! 1$. 
Furthermore, if the states $|F_{e},\,m\rangle$ are subject to spontaneous emission, the `dark spin-cat' is also autonomously stabilized.   

We first analyze the potential for longitudinal field perturbations ($2F_g\!- n_z\!\gg\!1$) to cause bit flip errors, which will a priori only occur when the two SCSs are not perfectly orthogonal, as \emph{e.g.} due to imperfect laser control. For the case of a small error $\epsilon$ in the angles $\theta_{1,2}, \varphi_{1,2}$, the off-diagonal matrix elements of perturbations $(\hat{F}_{g,z})^{n_z}$ are exponentially suppressed in $F_g$,
\begin{align}
|\langle \theta_1, \varphi_1|&(\hat{F}_{g,z})^{n_z}|\theta_2, \varphi_2 \rangle| = |\epsilon|^{2F_g-n_z} \times \\
&2^{-2F_g} \frac{(2F_g)!}{(2F_g-n_z)!} |\sin(\theta_1)|^{k}+\mathcal{O}\left(F_g^{n_z}\epsilon^{2F_g+1-n_z}\right)
.\notag
\end{align}
Here, $k\!=\!n_z$ for $(\theta_2,\, \varphi_2) = (\pi\!-\!\theta_1\!+\!\epsilon,\,\pi\!+\!
\varphi_1)$ and $k\!=\!2F_g$ for $(\theta_2,\, \varphi_2) = (\pi\!-\!\theta_1,\,\pi\!+\!
\varphi_1\!+\!\epsilon)$.
This exponential suppression of bit-flip transitions is the basic building block of the biased error model.

Moreover, keeping the SCSs close to the equatorial plane makes them also robust against dephasing. This is evident from 
the differential of the diagonal matrix elements in the logical qubit subspace \cite{aharon2013general}
\begin{align}
\langle \theta_1, \varphi_1|(\hat{F}_{g,z})^{n_z}&|\theta_1, \varphi_1 \rangle - \langle \theta_2, \varphi_2|(\hat{F}_{g,z})^{n_z}|\theta_2, \varphi_2 \rangle = \\
&= 2(F_g\cos\theta_1)^{n_z}+\cos\theta_1\,\mathcal{O}(F_g^{n_z-1})\notag
\end{align}
for orthogonal SCS $(\theta_2,\, \varphi_2) = (\pi-\theta_1,\,\pi+
\varphi_1)$, which vanishes for $\theta_1 = \pi / 2$.

\begin{figure}[t!]
\center
\includegraphics[width=0.99\columnwidth]{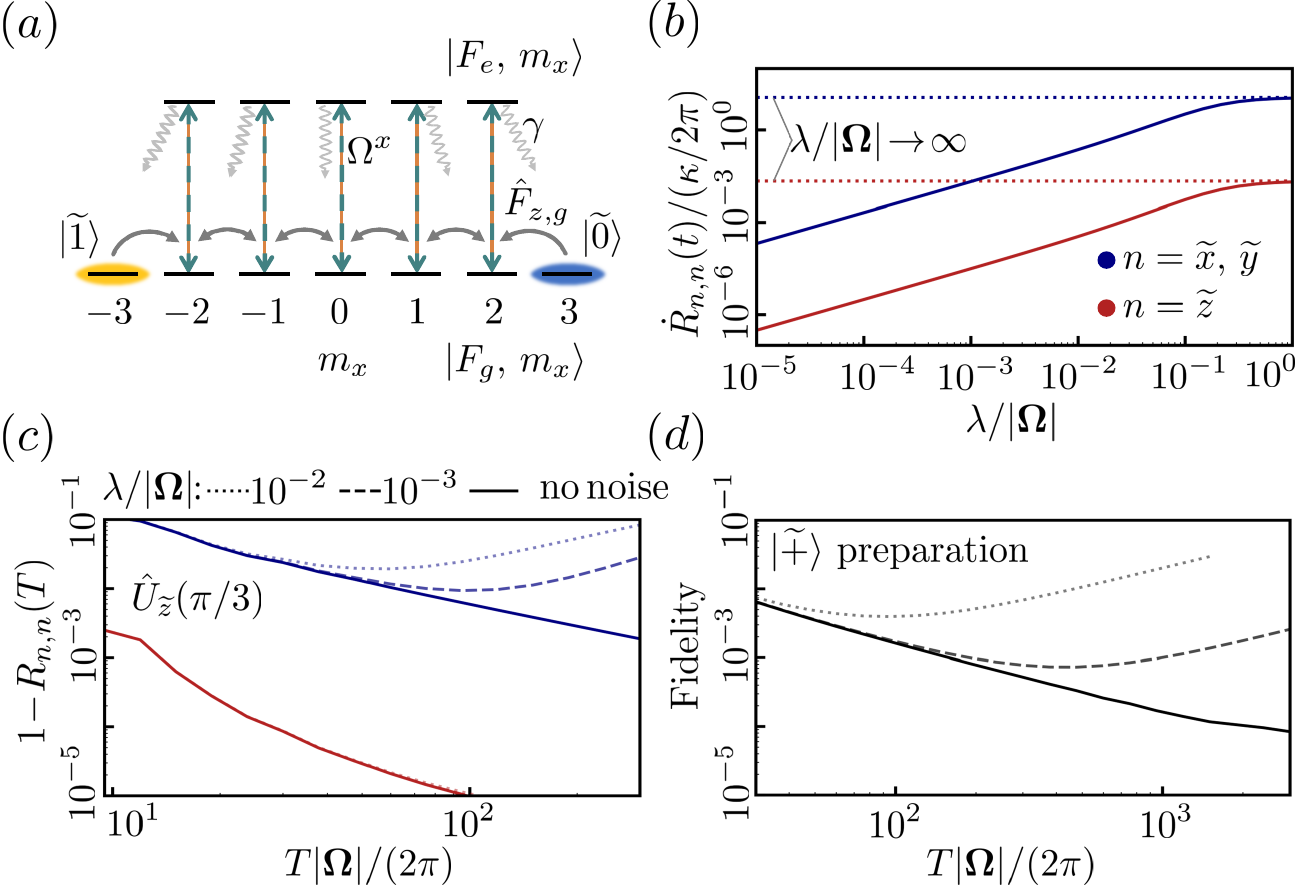}
\caption{(a) Autonomous stabilization: laser coupling (\mbox{$F_g\!=\!3$}) in the rotated coordinate system with quantization axis \mbox{$\boldsymbol{\Omega}=(\Omega^x,\,0,\,0)^T$}. 
Note that $\Delta\!=\!\delta\!=\! 0$ and $|F_{g/e},\,m_x\rangle$ are the eigenstates of $\hat{F}_{x,g/e}$. 
A noise process $\hat{F}_{z,g}$ couples states $|F_{g},\,m_x\rangle$ adjacent in $m_x$.
Decay of $|F_{e},\,m_x\rangle$ is due to the CG coefficients directed "towards" the closest DS, as illustrated by wavy arrows indicating the average $m_x$ change.
(b) Time derivative of the diagonal elements of the PTM $\mathbf{R}(t)$ (effective error rates) for different values of $\lambda/|\boldsymbol{\Omega}|$.
The time derivative is taken in the linear regime $t\kappa/(2\pi)\!\sim\!1$.
This and the remaining panels all share $\Delta\!=\!\delta=0$, $|\boldsymbol{\Omega}|/\gamma\!=\!2\pi$, $F_g\!=\!4$, $\kappa\!=\!10^{-4}|\boldsymbol{\Omega}|$ and $N_X\!=\!3$.
Panel (c) displays the error channels for $\hat{U}_{\widetilde{z}}(\alpha\!=\!\pi/3)$ and 
(d) shows the state  $|\widetilde{+}\rangle$ state preparation protocol fidelity for the same noise parameters as in (c).
}
\label{fig:3}
\end{figure}

\textit{Autonomous stabilization.-}
Another type of error that can occur is leakage from the logical qubit space. 
Such leakage errors can be converted primarily into dephasing errors by combining laser driving and spontaneous emission.
It is useful to first consider a frame transformation, by rotating the quantization axis to that determined by $\boldsymbol{\Omega}$, see SM\,\cite{SupMat}. In this frame, the laser drive is linearly polarized, coupling only the interior levels ($|m| < F_g$) to the $F_e$-manifold, see Fig.\,\ref{fig:3}(a). In this eigenbasis, the DSs are the two maximally stretched spin states of the $F_g$-manifold, separated by $2F_g - 1$ pairs of bright states. Noise processes associated to $(\hat{F}_{z,g})^{n_z}$, with $n_z\!\ll\! F_g$, couple the DSs to the interior bright states. This leakage is determined by the noise strength relative to the gap from the dark to the relevant bright state, thus can be minimized by increasing $|\mathbf{\Omega}|$. 

Importantly, spontaneous emission, of strength $\gamma$, from the $F_e$ component of the bright states favors decay processes towards the closest DS due to the associated CG coefficients~\cite{sobel2016introduction}, effectively driving leaked population directly back to the qubit state it originated from, converting the error into dephasing.
Hence, the combined action of the laser driving and spontaneous emission is a bias-preserving stabilization process, in analogy to bosonic systems~\cite{guillaud2023quantum}. 
The time scale of autonomous stabilization is in the limit $|\mathbf{\Omega}|\!\gg\!\gamma$ determined by $\gamma/(2F_g+1)$ as discussed in the SM\,\cite{SupMat}.
Autonomous stabilization is particularly suited to broad $F\!\leftrightarrow\!F\!-\!1$ laser-cooling transitions, with $F\!\gg\!1$, \emph{e.g.}, the ground state of Nd or meta-stable excited-states of Dy \cite{hovhannesyan2023improving}. 
Further, the amplitude error mentioned in~\cite{omanakuttan2024fault} can be corrected passively given that we can engineer specific polarizations of the emission~\cite{pineiro_orioli_emergent_2022}, see SM~\cite{SupMat}.

\textit{Colored noise.-} 
We now analyze the robustness of the `dark spin-cat' subjected to colored Markovian noise processes $X(t)$, governed by an Ornstein-Uhlenbeck process, with correlation time $1/\lambda$ and diffusion constant $\lambda^2\kappa/2$, coupling to $\hat{F}_{g,z}$, \emph{e.g.} a fluctuating magnetic field. 
Note, $\kappa$ is the noise strength in the white noise limit $\lambda/|\boldsymbol{\Omega}|\!\rightarrow \!\infty$ \cite{gardiner2009stochastic}.
Following Miao~\cite{miao2013analysis}, we directly compute the stochastic average of the system state $\hat{\rho}(t)$ over all possible noise trajectories~\cite{van1976stochastic, zoller1981ac, SupMat}.
The resulting error channel can be characterized by the Pauli transfer matrix (PTM)~\cite{nielsen2010quantum, greenbaum2015introduction}
\begin{align}
R_{n,m}(t) = \text{tr}\left[\hat{E}_n\,\hat{\rho}_m(t) \right]/2 \text{ for } \hat{\rho}_m(0) =\hat{E}_{m},
\end{align}
with $\hat{E}_n\in \{\widetilde{\mathbb{1}},\,\hat{\sigma}_{\widetilde{x}},\,\hat{\sigma}_{\widetilde{y}},\,\hat{\sigma}_{\widetilde{z}}\}$, where $\widetilde{\mathbb{1}}$ denotes the qubit subspace projector.
The diagonal elements of $\mathbf{R}(t)$ carry the relevant error channel information \cite{erhard2019characterizing, magesan2011scalable}, where   $R_{\widetilde{x},\widetilde{x}}(t)\!=\!R_{\widetilde{y},\widetilde{y}}(t)$ and $R_{\widetilde{z},\widetilde{z}}(t)$ are associated with the dephasing and bit-flip error, respectively, see SM\,\cite{SupMat}. 

In Figs.\,\ref{fig:1}(c) and \ref{fig:3}(b) we present the diagonal elements of $\mathbf{R}(t)$
for `dark spin-cat' qubits subjected to colored noise, exhibiting a suppression of both bit-flip and dephasing errors by increasing laser coupling strength. 
Bit-flip error rates decrease exponentially in $F_g$, while phase-flip errors increase only polynomially.
Note, in Fig.\,\ref{fig:1}(c) the same parameters as in \ref{fig:3}(b) are used.
Typical values of $\boldsymbol{\Omega}$ for electric dipole transitions are several hundred MHz, for which we observe a severe reduction of effective noise rates, even for very strong noise with $\kappa, \lambda\!\sim\! O(10\!-\!100 \mathrm{kHz}$).
The overall strength of the error rates is determined by the power spectral density at the gap frequency $\propto |\boldsymbol{\Omega}|$.
For $\lambda \gtrsim |\boldsymbol{\Omega}|$, the noise becomes white and is described by Lindbladian dynamics with jump-operator $\sqrt{\kappa}\hat{F}_{g,z}$ [see Fig.~\ref{fig:3}(b)], for which we still observe an exponential bias \cite{SupMat}.

\textit{Logical operations.-} 
We now discuss error channels of single-qubit gates in the presence of colored noise processes as discussed above.
In Fig.\,\ref{fig:3}(c) we present for $\hat{U}_{\widetilde{z}}(\alpha)$ the diagonal elements of $\mathbf{R}(T)$ adjusted by the inverse error-free unitary gate.
The worst-case gate infidelity is, due to the biasedness of the channel \mbox{$1\!-\!R_{\widetilde{z},\widetilde{z}}(T)\ll1\!-\!R_{\widetilde{x},\widetilde{x}}(T)$}, given by $1\!-\! \mathcal{F} = [1\!-\!R_{\widetilde{x},\widetilde{x}}(T)]/2$\,
\footnote{
States most sensitive to dephasing are on the equator of the logical qubit Bloch-sphere. 
The worst case fidelity is obtained by uniformly sampling over these states.}.
At small $T$, the gate fidelity is limited by adiabaticity violations, while at large $T$ the noise $X(t)$ is the limiting factor. 
The former can be mitigated with counter-diabatic driving techniques~\cite{del2013shortcuts, SupMat}.
Nevertheless, for $|\boldsymbol{\Omega}|\!=\!2\pi\times300\,\mathrm{MHz}$ the gate can be executed as fast as $T\!=\!1\,\mu s$ in a bias preserving manner, with an infidelity below $10^{-2}$.
We note that $\hat{U}_{\widetilde{x}}$ can be performed \textit{virtually}, with perfect fidelity, by swapping definitions of $\ket{\widetilde{0}}$ and $\ket{\widetilde{1}}$.

\textit{Initialization.-} Preparation of a logical state, such as \mbox{$\ket{\widetilde{+}}\! =\! (\ket{\widetilde{0}} +  \ket{\widetilde{1}})/\sqrt{2}$}, can be accomplished by first preparing the qubit in the stretched state $|F_g, -F_g \rangle$ (for $\Omega^{-1}\! >\! 0$) using optical pumping. 
This state is then adiabatically converted by first ramping on $\Omega^{+1}$, akin to a STIRAP protocol \cite{vitanov2017stimulated} (see SM\,\cite{SupMat}). The corresponding fidelity, \emph{i.e.} overlap, is presented in Fig.\,\ref{fig:3}(d). Measurement can be performed by reversing the preparation method and subsequently monitoring the population of $\ket{F_g,-F_g}$.

\begin{figure}[t]
\center
    \includegraphics[width=0.99\columnwidth]{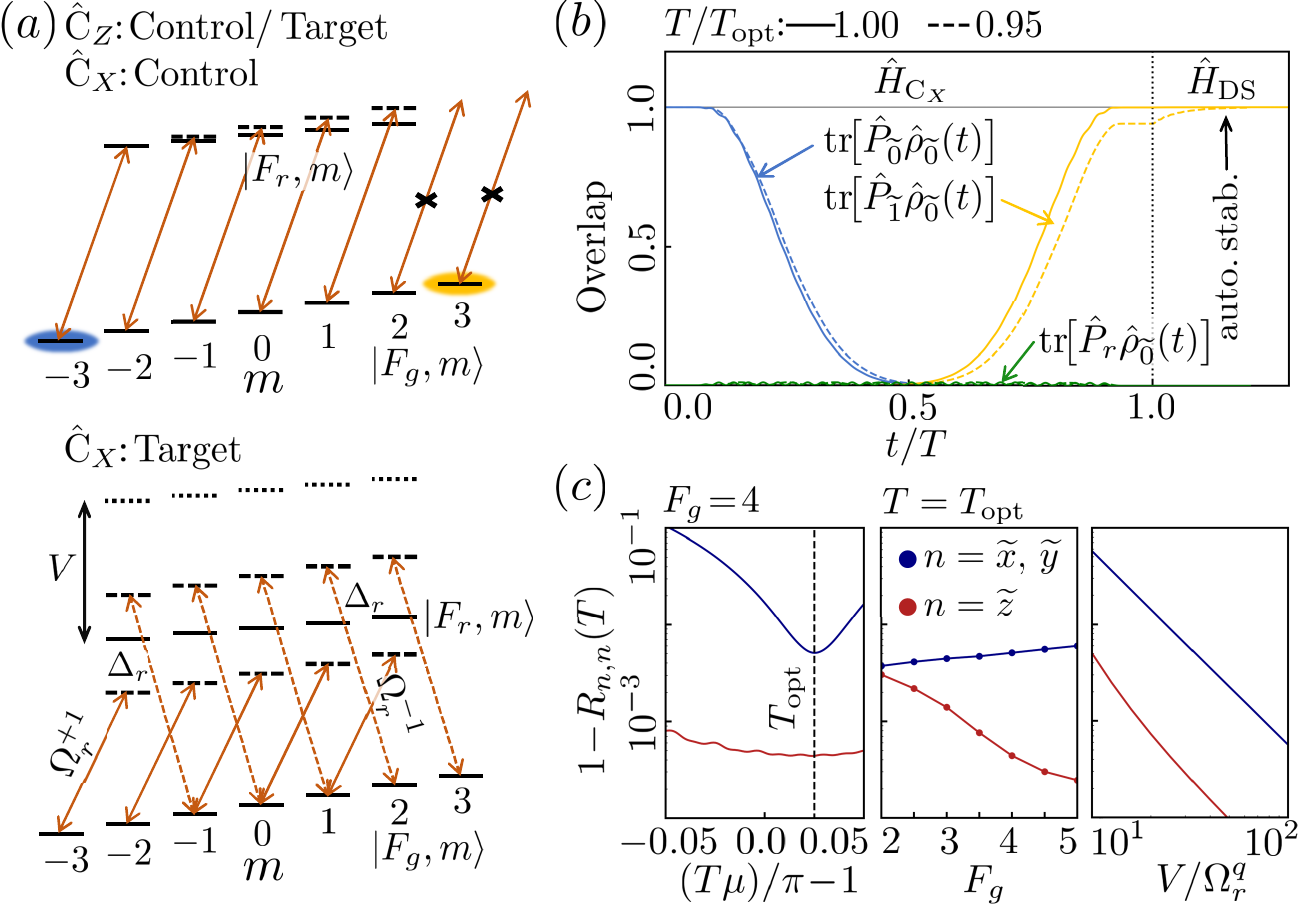}
\caption{Entangling gates: 
(a) $\hat{\mathrm{C}}_Z$ and $\hat{\mathrm{C}}_X$ laser coupling scheme to Ry states for \mbox{$F_g\!=\!3$}. 
For the $\hat{\mathrm{C}}_Z$-gate execution the target and control atom are encoded along the $z$-axis, while for the $\hat{\mathrm{C}}_X$-gate the control atom is encoded along the $z$-axis and the target atom along the $x$-axis.
(b) $\hat{\mathrm{C}}_X$: Time evolution of the target atom starting from $|\widetilde{0}\rangle$, when the control atom is not Ry excited, for two different gate times. $\hat{P}_{\widetilde{0}}$, $\hat{P}_{\widetilde{1}}$, $\hat{P}_{r}$ denote projectors onto $|\widetilde{0}\rangle$, $|\widetilde{1}\rangle$ and all $|F_r,\, m\rangle$ states, respectively. 
Here, and in all remaining panels, $F_g\!=\!4$, $\Omega_r^q/\Delta_r\!=\!2$, $\Omega^q_r\!=\!2\pi\!\times\!3\,\mathrm{MHz}$ and $\delta_r\!=\!0$ if not stated otherwise.
(c) $\hat{\mathrm{C}}_X$: Diagonal elements of $\mathbf{R}(T)$ for different gate times (left), at the optimal gate time $T_\mr{opt}$ for different $F_g$ (middle) and when the control atom is Ry excited for different $V$ (right).
}
\label{fig:4}
\end{figure}

\textit{Entangling gates.-} 
A universal set of quantum gates requires the implementation of an entangling operation, which we exemplify here for neutral atoms laser-excited to Ry states. We consider `dark spin-cat' qubits encoded in the maximally stretched HF manifold of the meta-stable ${}^3P_2$~\cite{klusener2024coherent} fine structure manifold of fermionic divalent atoms, such as \textsuperscript{171}Yb, \textsuperscript{173}Yb or \textsuperscript{87}Sr, for which $F_g\! =\! 5/2,\,9/2,\,13/2$, respectively.
Autonomous stabilization is achieved by coupling to the maximally stretched lowest ${}^3S_1$-manifold
\footnote{We note that this transition is only partially cyclic, thus spontaneous emission leads to decay outside the $F_g$-manifold, requiring a suitable choice of repumping lasers.}.


First, we outline the implementation of a $\hat{\mathrm{C}}_Z$-gate, which, together with \emph{e.g.} $\hat{U}_{\widetilde{z}}(\alpha_z)$ and $\hat{U}_{\widetilde{x}}(\alpha_x)$ forms a universal gate set. 
Detailed information on \(\hat{U}_{\widetilde{x}}(\alpha_x)\) is provided in the SM~\cite{SupMat}.
For the $\hat{\mathrm{C}}_Z$ gate, the control and target atom are rotated to SCS pointing along the $z$-axis enabling selective excitation to a maximally stretched ${}^3S_1$ ($F_r\!=\! F_g\! - 1$) Ry state \cite{Robicheaux_2019}, see Fig.\,\ref{fig:4}(a), which allows for the execution of state-of-the-art $\hat{\mathrm{C}}_Z$-gates~\cite{evered2023high, ma2023high, cao2024multi}.
Additionally, neutral-atom platforms are amenable to erasure conversion techniques, \cite{wu2022erasure, ma2023high}. 
Especially the ${}^3P_2$ `dark spin-cat' encoding offers the possibility to pump leaked population to ${}^1S_0$ from where it can be detected \cite{yamamoto2016ytterbium, saskin2019narrow}.
For the $\hat{\mathrm{C}}_Z$-gate, only one of the two qubit states is Ry excited. Leakage due to spontaneous emission thus originates predominantly only from one of the qubit states, enabling high-threshold QEC strategies based on \textit{biased} erasure~\cite{sahay2023high}.
Additionally, Ry decay events that do return to the $F_g$-manifold do not introduce bit-flip errors due to autonomous stabilization. Thus, the `dark spin-cat' encoding naturally realizes biased erasure conversion alongside a reduced and structured error model, which has the potential to further enhance the error threshold.

To harness full advantage of the `dark spin-cat's biased noise structure, bias preserving QEC codes have to be employed. 
Such codes require the direct execution of biased $\hat{\mathrm{C}}_X$ (CNOT) and $\hat{\mathrm{CC}}_X$ (Toffoli) gates~\cite{guillaud2019repetition}. We provide a detailed protocol for the $\hat{\mathrm{C}}_X$ gate in the following. 
The implementation is based on an artificial local magnetic field $\mu \hat{F}_z$ ($\mu$ is the field strength), engineered through light-induced AC Stark shifts \cite{Cohen1972, le2013dynamical}, exchanging the qubit states of the target atom.
Its action can be made conditional on the state of the control atom by means of the Ry blockade effect \cite{Jaksch2000, Lukin2001}.
The target qubit is encoded in SCSs along the $x$-axis with the stabilization turned off during the gate. 
The artificial magnetic field responsible for exchanging the qubit states is implemented by off-resonantly coupling the states $|F_g,\, m\rangle$ by two circularly polarised laser fields to the Ry-states mentioned for $\hat{\mathrm{C}}_Z$ labelled by $|F_r,\, m\rangle$.
The corresponding Hamiltonian is given by Eq.\,\eqref{eq:DSHamiltonian}, but with the replacement $e\! \leftrightarrow \! r$, $\Delta\! =\! 0$, and not on Raman resonance, \emph{i.e.} with couplings $\Omega_r^q\hat{\mathcal{C}}_q \exp(iq\Delta_r)$, see Fig.\,\ref{fig:4}(a).
In the limit \mbox{$|\Omega_r^{+1}|\!=\!|\Omega_r^{-1}|\!\ll\! |\Delta_r|$}, the laser coupling gives rise to an artificial magnetic field with strength 
$\mu = \frac{|\Omega^q_{r}|^2}{4\Delta_r}\frac{1}{F_g}\frac{2 F_g - 1}{2 F_g + 1}$
up to second order in $\Omega^q_{r}/\Delta_r$.

Similar to the $\hat{\mathrm{C}}_Z$-gate the control atom is rotated to the $z$-axis from where state selective Ry excitation is possible (Fig.\,\ref{fig:4}(a)).
Notably, the transformation to the $z$-axis is achieved by rotating the SCSs around the $y$-axis, which is in analogy to $\hat{U}_{\widetilde{x}}$ a bias-preserving process.
If excited, the control atom shifts the energies of the target atom due to Van der Waals interactions \cite{Robicheaux_2019}, with an interaction shift \mbox{$V\!\gg\! \Delta_r$}, rendering the artificial magnetic field inoperative. 
The resulting Hamiltonian reads \mbox{$\hat{H}_{\mathrm{C}_X}\!=\! \big(\mathbb{1}^{}\! - \!\hat{P}_r\big)\!\otimes\!\mu\, \hat{F}_{z,g}$} up to second order in $\Omega_r^q/\Delta_r$ and first order in $\Omega_r^q/V$, where $\hat{P}_r$ denotes the Ry state projector. 
The desired entangling gate is then effected by \mbox{$\hat{\mathrm{C}}_X\!=\! \exp\big(\!-\!i\hat{H}_{\mathrm{C}_X} \Tgate\big)$}, where $\Tgate\! \sim\! \pi / \mu$ is the gate time obtained from numerical simulations. 

In Fig.\,\ref{fig:4}(b) we present the time evolution of the target atom if the control atom is not Ry excited, demonstrating a swap of the qubit states. We also analyse a deliberate under-rotation, inducing leakage from the qubit subspace. This population can be re-pumped by turning on the autonomous stabilisation, $\hat{H}_\mathrm{DS}$ from Eq.\,\eqref{eq:DSHamiltonian}. Overall, this leads to a bias preserving gate, even in the presence of over- or under-rotation errors (see left and middle panel Fig.\,\ref{fig:4}(c)).

In the middle panel of Fig.\,\ref{fig:4}(c) we present dephasing and bit-flip errors when the control atom is Ry excited.
The bit-flip error is exponentially suppressed in $F_g$ as compared to the dephasing error. 
Deviations from exponential suppression for large $F_g$ are due to higher order corrections and can be reduced by decreasing $\Omega^q_r/\Delta_r$. In the right panel of Fig.~\ref{fig:4}(c) we show the same errors as a function of the Rydberg blockade strength $V$.
The overall worst-case gate infidelity $1\!-\!\mathcal{F}\!=\! [1\!-\!R_{\widetilde{x},\widetilde{x}}(T)]/2$ (as discussed above) can be on the order of $10^{-3}$, with \mbox{$\Omega^q_r=2\pi\!\times\!3\,\mathrm{MHz}$}, \mbox{$\Delta_r=2\pi\!\times\!6\,\mathrm{MHz}$} resulting in \mbox{$T_\mathrm{opt}=8.5\mu s$} requiring $V\!\sim\!2\pi\!\times\!O(100\,\mathrm{MHz})$ for $F_g\!=\!4$. 
For this analysis we neglected decay of the Ry excited control atom, which would in fact lead to non-bias preserving processes. 
These processes can be circumvented by multiplying the number of control atoms \cite{Cong2022} or detecting the decay of a control atom using erasure conversion \cite{ma2023high}. A more thorough analysis incorporating control atom decay is provided in the SM\,\cite{SupMat}. Note that a $\hat{\mathrm{CC}}_X$ gate can also be implemented with our protocol by separately exciting two control atoms instead of one~\cite{SupMat}.

\textit{Conclusions.-} 
We introduced an atomic `dark spin-cat' qubit encoding featuring a bias-preserving error model and autonomous stabilization.
We provide proof-of-principle gate implementations on a Ry platform, but emphasize that all results
can not only be improved significantly using optimal control techniques or counter-adiabatic driving~\cite{berry2009transitionless,del2013shortcuts} but also directly extended to other platforms such as trapped ions \cite{mikelsons2015universal}.
In particular, the single-qubit control can be extended to trapped ions and an entangling gate is realizable by rotating the qubits to the z-axis from where the geometric phase entangling gate~\cite{leibfried2003experimental} is executable; however, we reserve detailed analysis for future work.
The encoding scheme could also benefit quantum simulations with magnetic atoms ${}^{167}\mathrm{Er}$ or ${}^{161}\mathrm{Dy}$ and hetero-nuclear molecules ${}^{40}\mathrm{K}{}^{87}\mathrm{Rb}$, engineering a closed qubit subspace with strong and tunable dipolar interactions~\cite{perlin2022spinModels, kruckenhauser2020quantum}. 
Higher order multipole couplings enable the construction of `multi-legged' spin-cats, offering the possibility to redundantly encode quantum information in a single atom in order to tolerate more errors while preserving large error-bias~\cite{leghtas2013hardware, mirrahimi2014dynamically}.
Overall, spin-cat encodings with biased error models implemented in single atoms could provide a viable platform enabling resource-efficient quantum error correction codes, meeting the demanding requirements of fault-tolerant quantum computation.

\textit{Acknowledgements.-} 
We would like to thank J. Zeiher, M. Ammenwerth, G. Giudice, A. Retzker, I. Deutsch, M. Lukin, A. Kaufman, J. Ye and A. M. Rey for enlightening discussions.
This work was supported by the European High-Performance Computing Joint Undertaking (JU) under grant agreement No 101018180 HPCQS, the Horizon Europe programme HORIZON-CL4-2022-QUANTUM-02-SGA via the project 101113690 (PASQuanS2.1), the ARO(W911NF-23-1-0077), ARO MURI (W911NF-21-1-0325), AFOSR MURI (FA9550-19-1-0399, FA9550-21-1-0209, FA9550-23-1-0338), DARPA (HR0011-24-9-0359, HR0011-24-9-0361), NSF (OMA-1936118, ERC-1941583, OMA-2137642, OSI-2326767, CCF-2312755), NTT Research, Samsung GRO,  Packard Foundation (2020-71479). 

\newpage

\onecolumngrid

\appendix

\section*{Appendix}

\section{Spin coherent states as the dark states of the driving Hamiltonian}
\label{A}

In this Section we review the derivation of the system Hamiltonian from Eq.\,(1) and provide a detailed derivation of the system dark sates (DSs). 

\subsection{System Hamiltonian}

Here we review the derivation of the system Hamiltonian. 
The system we consider in the main-text is composed of two light coupled spin manifolds of length $F_g$ and $F_e=F_g-1$, respectively. 
The corresponding spin states are denoted by $|F_g,\,m\rangle$ and $|F_e,\,m\rangle$. 
The atomic level structure, see Fig.\,1(e), is given by an overall manifold splitting $\omega_{eg}$, with further magnetic substructure.
In the presence of a static magnetic field $\mathbf{B}$ (defining the quantisation axis) the magnetic sublevels within a spin-manifold of states are split by the Zeeman shift $\delta_{g/e} = g_{g/e}\mu_B|\mathbf{B}|$, where $g_{g/e}$ is the manifold specific Land\'e-g factor and $\mu_B$ the Bohr magneton. 
The system Hamiltonian is given in the laboratory frame by
\begin{align}
\hat{H}^\mathrm{lab}_\mathrm{DS}/\hbar = \omega_{eg} \hat{P}_e + \delta_e \hat{F}_{e,z} + \delta_g \hat{F}_{g,z} +\frac{1}{2}\sum_{q = 0,\pm 1} \left(e^{-i\omega_qt} \Omega^q \hat{\mathcal{C}}_q + \mathrm{h.c.}\right),
\end{align}
where the operator $\hat{P}_e = \sum_m|F_e, m\rangle \langle F_e, m |$ is a projector onto the excited manifold of states, and where $\omega_q$ and $\Omega^q$ denotes the oscillation and Rabi frequency of the light field  with polarization $q\in\{\pm1,\,0\}$, respectively.  
Note, the Rabi frequency $\Omega^q$ is a contravariant vector $\Omega^q = (\Omega_{q})^*$, where $\Omega_q$ is the corresponding covariant vector \cite{king2008angular}, and the total Rabi frequency is given by $|\mathbf{\Omega}| = \sqrt{\sum_q \Omega^q\Omega_q}$.
Furthermore, the coupling between the two spin manifolds is given by the operator $\hat{\mathcal{C}}_q = \sum_m\mathcal{C}_{F_g,m; 1,q}^{F_e, m+q}|F_e,m+q\rangle \langle F_g, m |$, where 
\begin{align}
\label{SM:CG}
\mathcal{C}_{F_g,m; 1,q}^{F_e, m+q}&= \langle F_g,m;1,q\ket{F_e,m+q} = \sqrt{2 F_e+1}\frac{\langle F_e,m+q | \hat{T}^{(1)}_q |F_g, m \rangle}{\langle F_e || \hat{T}^{(1)} ||F_g \rangle} 
\end{align}
is a Clebsch-Gordan coefficient $\langle F_g,m_F;1,q\ket{F_e,m_F+q}$, which, according to the Wigner-Eckart Theorem \cite{sakurai1995modern}, is given by the transition matrix element of a spherical tensor $\hat{T}_q^{(k)}$ of rank $k=1$ (due to the electric/magnetic dipole coupling) and the corresponding reduced matrix element $\langle F_g||\hat{T}^{(1)}||F_e\rangle$.

If the oscillation frequencies of the three light fields fulfill the Raman resonance criteria $\omega_q - \omega_{q'} = \delta_g (q-q')$, there exists a rotating frame transformation 
\begin{align}
\hat{\mathcal{U}} = \exp\left\{i\left[\omega_0 \hat{P}_e + \delta_g(\hat{F}_{e,z} + \hat{F}_{g,z})  \right] t \right\}
\end{align}
for which $\hat{H}^\mathrm{lab}_\mathrm{DS}$ becomes time-independent
\begin{align}
\hat{H}_\mathrm{DS} &= \hat{\mathcal{U}}\,\hat{H}_\mathrm{DS}^\mathrm{lab}\, \hat{\mathcal{U}}^\dagger-i\hbar \,\hat{\mathcal{U}}\big(\partial_t\hat{\mathcal{U}}^\dagger\big) \notag\\
  &= -(\hbar\Delta \hat{P}_e + \hbar\delta \hat{F}_{e,z}) + \frac{1}{2}\sum_{q = 0,\pm 1} \left(\hbar\,\Omega^q \hat{\mathcal{C}}_q + \mathrm{h.c.}\right),
  \label{SM:H}
\end{align}
with the detunings $\delta = \delta_g - \delta_e$ and $\Delta = \omega_0-\omega_{eg}$.
This Hamiltonian is equivalent to the Hamiltonian from the main-text Eq.\,(1).

\subsection{Dark states}
\label{sec:DS}
In the following we present a method to find the two DSs $|\mathrm{DS}_{1,2}\rangle$ of $\hat{H}_\mathrm{DS}$ from Eq.\,\eqref{SM:H}, which satisfy \mbox{$\hat{H}_\mathrm{DS}|\mathrm{DS}_{1,2}\rangle=0$}. 
This DSs are fully encoded in the $F_g$-manifold of states and are thus insensitive to spontaneous emission from the excited states and also to the precises values of $\delta$ and $\Delta$.
If $|\mathbf{\Omega}|\neq0$ the number of DSs is always given by $2$ \cite{morris1983reduction}.
A result of this method is that the two DSs of $\hat{H}_\mathrm{DS}$ live in a Hilbert space spanned by two spin coherent states (SCSs) (up to a singular hyper parameter space), which are not necessarily orthogonal.

\begin{figure}[!t]
\center
    \includegraphics[width=1\columnwidth]{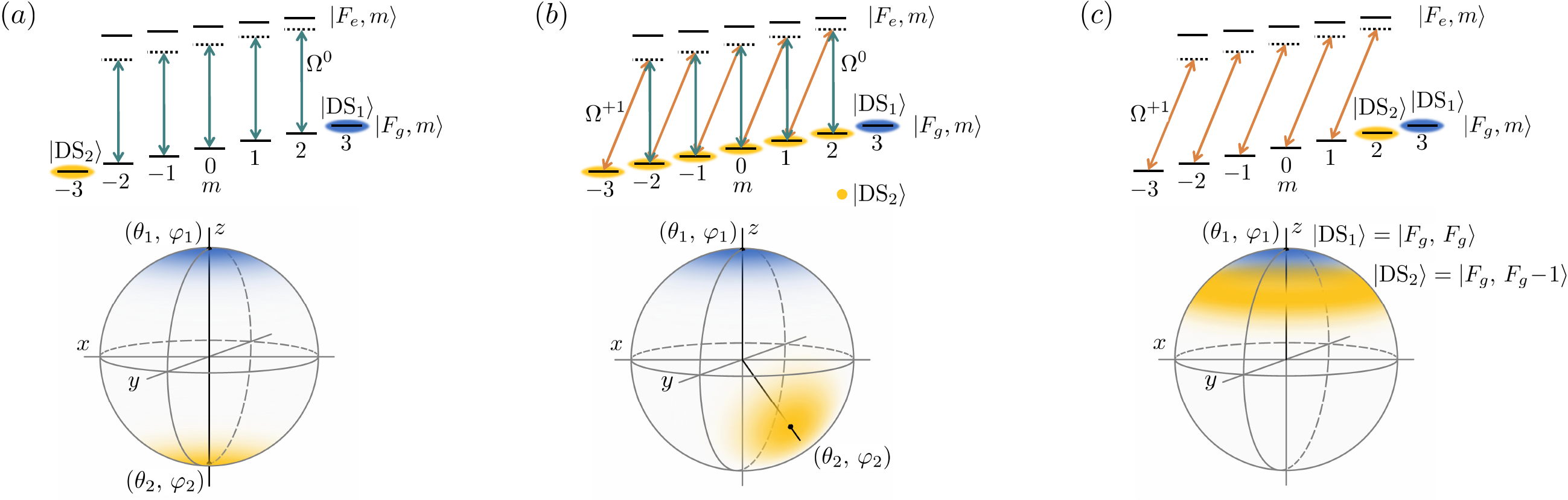}
\caption{Panels (a) - (c) Light coupling schemes, DSs and Wigner distributions for the three different examples discussed in the appendix section \ref{sec:DS} for $F_g=3$.}
\label{fig:SM1}
\end{figure}

From a pedagogical point of view it is useful to first discuss three simple special cases and then describe the more general method.
\begin{enumerate}
\item
We first focus on the case $\Omega^{\pm 1} = 0$ and $\Omega^0 = \Omega$,  
for which the two maximally stretched states $|\mathrm{DS}_{1,2}\rangle = |F_g,\,m=\pm F_g \rangle$ are not coupled by the light field and are thus the DSs of the system, see Fig.\,\ref{fig:SM1}(a). 
Furthermore, the two maximally stretched states $|F_g,\,F_g \rangle = |\theta = 0,\,\varphi = 0 \rangle$ and  $|F_g,\,-F_g \rangle = |\theta = \pi,\,\varphi = 0 \rangle$ are SCSs pointing along the $\pm z$-axis of the generalised $F_g$ Bloch sphere.
Hence, the DSs are given by two orthogonal SCSs.

\item 
Secondly, we consider the situation $\Omega^{- 1} = 0$ and $\Omega^0 = \Omega^{+1}= \Omega/\sqrt{2}$,
for which only the maximally stretched state $|\mathrm{DS}_{1}\rangle = |F_g,\,F_g \rangle $ is uncoupled and, therefore, is one of the two DSs, see Fig.\,\ref{fig:SM1}(b).
$|\mathrm{DS}_{2}\rangle$ is a superposition of states $|F_g,\,m\rangle$, with $m<F_g$.
We will show below that $|\mathrm{DS}_{2}\rangle$ is given by a SCS, which has to be Schmidt orthogonlized with respect to $|\mathrm{DS}_{1}\rangle$. 
Thus, the Hilbert space spanned by $|\mathrm{DS}_{1}\rangle$ and $|\mathrm{DS}_{2}\rangle$ is equivalent to the one spanned by two non-orthogonal SCSs.

\item 
An example of the above mentioned single hyper parameter space is $\Omega^{+1} = \Omega$ and $\Omega^{-1} = \Omega^{0} =  0$. Here the DSs can not be expressed by two SCSs as illustrated in Fig.\,\ref{fig:SM1}(c). Insted, they are given by
$|\mathrm{DS}_1\rangle=|F_g,\,F_g\rangle$ and $|\mathrm{DS}_2\rangle=|F_g,\,F_g - 1\rangle$, where only $|\mathrm{DS}_1\rangle$ is a SCS. 
\end{enumerate}

The method we present in the following is based on spatial rotations $\hat{\mathcal{R}}$ such that the vector of Rabi frequencies $\mathbf{\Omega} = (\Omega^{+1},\,\Omega^{0},\,\Omega^{-1})^\mathrm{T}$ in the rotated coordinate system $\mathbf{\Omega}\overset{\hat{\mathcal{R}}}{\rightarrow}\widetilde{\mathbf{\Omega}}$ obeys $\widetilde{\Omega}^{-1} = 0$.
If this condition is fulfilled we can identify one of the two DSs (see example 2 from above) as the maximally stretched state $|F_g,\,F_g \rangle_{\widetilde{z}}$ along the $\widetilde{z}$-axis of the transformed coordinate system. 
The second SCS can be obtained by another rotation $\hat{\mathcal{R}}'$ 
for which $\widetilde{\Omega}'{}^{-1}=0$ is satisfied and $|\mathrm{DS}_2\rangle$ can be obtained by a Schmidt orthogonalization with respect to $|\mathrm{DS}_1\rangle$.

To formalize this method we now discuss properties of $\hat{H}_\mathrm{DS}$ from Eq.\,\eqref{SM:H} under spatial rotations
\begin{align}
\hat{\mathcal{R}}(\alpha ,\beta ,\gamma )=e^{-i\alpha \left(\hat{F}_{g, z} + \hat{F}_{e, z}\right)}e^{-i\beta \left(\hat{F}_{g, y} + \hat{F}_{e, y}\right)}e^{-i\gamma \left(\hat{F}_{g, z} + \hat{F}_{e, z}\right)},
\end{align}
where $\alpha,\,\beta$ and $\gamma$ are extrinsic Euler angles. 
Under such a rotation $H_{\mathrm{DS}}$ becomes
\begin{align}
\hat{\mathcal{R}}^\dagger \hat{H}_\mathrm{DS}\hat{\mathcal{R}}\,\frac{\langle F_e||\hat{T}^{(1)} ||F_g \rangle}{\sqrt{2F_e+1}} \frac{1}{\hbar} \stackrel{\delta=\Delta=0}{=} \sum_{m, m'} \hat{\mathcal{R}}^\dagger |F_e, m'\rangle \langle F_e, m'|\hat{\mathcal{R}} \,\left(\sum_q \Omega^q \left[\hat{\mathcal{R}}^\dagger \hat{T}_q^{(1)} \hat{\mathcal{R}}\right]\right)\hat{\mathcal{R}}^\dagger|F_g, m \rangle \langle F_g, m|\hat{\mathcal{R}} + \mathrm{h.c.},
\end{align}
where we expressed the CL coefficients in terms of $\hat{T}_q^{(1)}$ using Eq.\,\eqref{SM:CG}.
Note, as the DSs are independent of $\delta$ and $\Delta$ we dropped the detuning terms in Eq.\,\eqref{SM:H} out of simplicity.
Exploiting the transformation properties of spherical tensor operators \cite{sakurai1995modern} the inner bracket of the above equation can be rewritten as
\begin{align}
\sum_q \Omega^q \left[\hat{\mathcal{R}}^\dagger \hat{T}_q^{(1)} \hat{\mathcal{R}}\right] = \sum_{q} \Omega^q \sum_{q'} D^{(1)*}_{q,q'}\, \hat{T}_{q'}^{(1)},
\end{align}
where $D^{(1)}$ is the Wigner D-matrix, whose elements are given   
\begin{align}
D^{(1)}_{q,q'} = e^{-iq\alpha} d^{(1)}_{q,q'} e^{-iq'\gamma}
\end{align}
in terms of the small Wigner d-matrix 
\begin{align}
d^{(1)} = 
\begin{pmatrix}
\frac{1+\cos\beta}{2} &- \frac{\sin\beta}{\sqrt{2}} & \frac{1-\cos\beta}{2}\\
\frac{\sin\beta}{\sqrt{2}} & \cos \beta & -\frac{\sin\beta}{\sqrt{2}}\\
\frac{1-\cos\beta}{2} &\frac{\sin\beta}{\sqrt{2}} & \frac{1+\cos\beta}{2}
\end{pmatrix}~
\begin{bmatrix}
+1\\
0\\
-1
\end{bmatrix}.
\end{align}
From the transformation properties of the spherical tensor operators we identify the transformed coupling strengths as
\begin{align}
\widetilde{{\Omega}}^{q'} = \sum_q {\Omega}^q D^{(1)*}_{q,q'}
\end{align}

As discussed above we are interested in rotations for which $\widetilde{\Omega}^{-1} = 0$.
The rotation angles can be extracted from the solution of the following trigonometric equation 
\begin{align}
\widetilde{\Omega}^{-1} = e^{i\alpha}\left(\frac{1-\cos \beta}{2} \right) \Omega^{+1} - \frac{\sin \beta}{\sqrt{2}} \Omega^{0} + e^{-i\alpha}\left(\frac{1+\cos \beta}{2} \right) \Omega^{-1} = 0,
\end{align}
where we dropped the $\gamma$ term as this only adds a global phase (for the remainder we take without loss of generality $\gamma = 0$).
In general, $\widetilde{\Omega}^{-1}=0$ for two distinct rotations $\hat{\mathcal{R}}(\alpha_{1,2},\beta_{1,2}, 0)$.
The SCSs spanning the DS subspace are given by 
\begin{align} \label{eq:SCS}
|\theta_{1,2} = \beta_{1,2},\,\varphi_{1,2} = \alpha_{1,2}\rangle = \hat{\mathcal{R}}(\alpha_{1,2},\beta_{1,2}, 0)|F_g,\,F_g\rangle.
\end{align}
Note, the definition here differs to the definition of Radcliffe \cite{radcliffe1971some} by $|\theta, \varphi\rangle = e^{-iF\varphi}|\theta, \varphi\rangle_\mathrm{Radcliffe}$.
Two orthogonal DSs can, for instance, be defined as 
\begin{align}
\label{eq:DSorthogonal}
|\mathrm{DS}_1\rangle &= |\theta_1,\,\varphi_1\rangle\notag\\
|\mathrm{DS}_2\rangle &= \left(|\theta_2,\,\varphi_2 \rangle - \langle \theta_1,\,\varphi_1  |\theta_2,\,\varphi_2 \rangle |\theta_1,\,\varphi_1 \rangle \right) / \sqrt{1 - | \langle \theta_1,\,\varphi_1  |\theta_2,\,\varphi_2 \rangle|^2},
\end{align}
where we Schmidt orthogonalized $|\mathrm{DS}_2\rangle$ with respect to  $|\mathrm{DS}_1\rangle$.

The method discussed above gives to orthogonal DSs only when the tuples of angles are distinct, \emph{i.e.} $(\theta_1,\,\varphi_1)\neq (\theta_2,\,\varphi_2)$.
For $(\theta_1,\,\varphi_1) = (\theta_2,\,\varphi_2)$ the Rabi frequencies $\widetilde{\Omega}^q$ in the rotated coordinate system not only give $\widetilde{\Omega}^{-1} = 0$ but also $\widetilde{\Omega}^{0} = 0$.
For this special case, see example 3 from above, the two states in the rotated coordinate system $|F_g,\,F_g\rangle_{\widetilde{z}}$ and  $|F_g,\,F_g - 1\rangle_{\widetilde{z}}$ are not coupled by the light field.
Hence, the two DSs can not be expressed by two SCS, and can instead be chosen as  $|\mathrm{DS}_1\rangle = \hat{\mathcal{R}}(\alpha_1,\beta_1,0)|F_g,\,F_g\rangle$ and $|\mathrm{DS}_2\rangle = \hat{\mathcal{R}}(\alpha_1,\beta_1,0)|F_g,\,F_g -1\rangle$.
Note, regardless of the specific vales of $(\theta_1,\,\varphi_1)$ and $(\theta_2,\,\varphi_2)$ it is always possible to define two DS with a well defined parity.
That is, one DS only occupies even and the other DS only odd $m$ Zeeman-levels, respectively.

Let us finally discuss the special situation where the two SCSs are orthogonal. 
This situation arises if the Cartesian components of the Rabi frequency (in the rotating frame defined by $\hat{\mathcal{U}}$) 
\begin{align}
\Omega^x &= \frac{1}{\sqrt{2}}(\Omega^{-1} - \Omega^{+1}),~
\Omega^y = \frac{i}{\sqrt{2}}(\Omega^{-1} + \Omega^{+1}),~\mathrm{and}~
\Omega^z = \Omega^0
\end{align}
are up to a global phase real numbers. 
Under these conditions the Cartesian Rabi frequency vector $\mathbf{\Omega} = \left(\Omega^x,\,\Omega^y,\,\Omega^z \right)$ defines the $\widetilde{z}$-axis of the rotated coordinate system.
Therefore, the rotation angles of $\hat{\mathcal{R}}(\alpha,\,\beta,\,0)$ are given by 
\begin{align}
\alpha &= \mathrm{sgn}(\Omega^y)\arccos\left[\frac{\Omega^x}{\sqrt{(\Omega^x)^2 + (\Omega^y)^2}}\right], ~\mathrm{and}\notag\\
\beta &= \arccos\left[\frac{\Omega^z}{\sqrt{(\Omega^x)^2 + (\Omega^y)^2 + (\Omega^z)^2}} \right]
\end{align}
for which $\widetilde{\Omega}^{\pm1} = 0$ and $\widetilde{\Omega}^0 = \Omega$
and the DSs can be identified as the maximally polarized states $|F_g,\,\pm F_g\rangle_{\widetilde{z}}$, as discussed in example 1 from above. 
A possible choice of DSs in this situation is $|\mathrm{DS}_1\rangle = \hat{\mathcal{R}}(\alpha,\,\beta,\,0)|F_g,\, F_g\rangle=|\beta,\alpha\rangle$ and $|\mathrm{DS}_2\rangle = \hat{\mathcal{R}}(\alpha,\,\beta,\,0)|F_g,\, -F_g\rangle=e^{-i\pi F_g}|\pi - \beta,\,\pi + \alpha\rangle$.

\section{Colored noise simulation}
\label{sec:Colored}

Here we discuss the numerical methods used for simulating the system dynamics under colored noise used in the main-text to generate Figs.\,1(c) and 3(a-d). 
For this, we consider a Markovian noise process $X(t)$ that couples to a system operator $\hat{\mathcal{O}}$, for example to $\hat{F}_{g,z}$ as considered in the main-text. 
The corresponding generalized multiplicative stochastic master equation is given by 
\begin{align}
\hbar\partial_t \hat{\rho}(t) &= \hat{\mathcal{L}}\hat{\rho}(t) -i\left[\hbar\,X(t) \hat{\mathcal{O}},\hat{\rho}(t)\right],\text{ with } \notag\\
    \hat{\mathcal{L}}\hat{\rho}(t) &= -i \left[\hat{H},\,\rho(t) \right] + \hbar \sum_a \gamma_a \mathcal{D}\left[\hat{a}\right] 
\label{eq:stochsticME}
\end{align}
where $\hat{\mathcal{L}}$ is the system Lindbladian decomposed into a Hamiltonian part $\hat{H}$ and dissipative part described by jump operators $\hat{a}$, with rate $\gamma_a$ and  $\mathcal{D}[\hat{a}]\hat{\rho}(t) = \hat{a}\hat{\rho}(t) \hat{a}^{\dagger} - \frac{1}{2} \{\hat{a}^{\dagger}\hat{a}, \hat{\rho}(t)\}$.
Importantly, $\hat{\rho}(t)$ denotes the density operator for a particular noise trajectory $X(t)$. 
The average density operator $\langle\hat{\rho}(t) \rangle_\mathrm{s}$, which is the quantity we wish to calculate, is defined by the statistical mean of $\hat{\rho}(t)$ over all possible noise trajectories $X(t)$. 
The noise trajectories are associated with a probability distribution $P(X,t)$, which satisfies a differential equation of the following form
\begin{align}
\hbar\partial_t P(X,t) = \hat{\Lambda}\,P(X,t),
\end{align}
where $\hat{\Lambda}$ is called Fokker-Planck (differential-) operator for continuous noise models, or, a transition rate matrix for jump-like noise models, respectively. 
In the main-text we considered a Ornstein-Uhlenbeck (OU) process which will be detailed below.

The statistical mean $\langle\hat{\rho}(t) \rangle_\mathrm{s}$ can either be calculated by trajectory sampling or alternatively directly employing the concept of marginal densities. 
In the following we will summarize the relevant formula using the latter approach, where we closely follow Ref.\,\cite{zoller1981ac}. 
The marginal density $\hat{u}(X, t)$, with elements $u_{\mu,\nu}(X, t)$, where $\mu,\nu$ are the indices corresponding to the density operator element $\rho_{\mu,\nu}(t)$, is related to the statistical mean of the density operator by \cite{van1976stochastic}
\begin{align}
\langle\hat{\rho}(t) \rangle_\mathrm{s} = \int \hat{u}(X,t)\,\mathrm{d}X.
\end{align}
The differential equation which $\hat{u}(X,t)$ satisfies reads
\begin{align}
\label{eq:margden}
\hbar\partial_t\kett{u(X,t)} = \left[\hat{\Lambda} -i\hbar X \hat{\hat{\mathcal{O}}} + \hat{\hat{\mathcal{L}}} \right] \kett{u(X,t)},
\end{align}
where $\kett{u(X,t)}$ is a column stacked vectorization of $\hat{u}(X,t)$, and $\hat{\hat{\mathcal{O}}}$ and $\hat{\hat{\mathcal{L}}}$ are the associated column stacked super-operators. 
The initial state of the marginal densities is given by $\hat{u}(X,t=0) = P_\mathrm{ss}(X)\,\hat{\rho}(0)$, where $P_\mathrm{ss}(X)$ is the steady-state noise distribution, which satisfies $\hat{\Lambda}P_\mathrm{ss}(X)=0$.

In the following we will provide explicit expressions of the super-operators for the colored noise simulations presented in the main-text. 
In the main-text we considered the OU process, whose Fokker-Planck operator is given by  
\begin{align}
\hat{\Lambda}/\hbar = \lambda\, \partial_X X  +\sigma^2/2\, \partial_X^2, 
\end{align}
where $1/\lambda$ and $\sigma^2$ determine the noise correlation time and diffusion constant, respectively \cite{gardiner2009stochastic}.
In the main-text we introduced $\kappa = \sigma^2/\lambda^2$, which corresponds to the strength of the noise in the white noise limit.
In particular, if $\lambda$ is much larger (smaller) than the characteristic energy scale of $\hat{H}$, the noise effectively appears white (static), as discussed below. 

The numerical integration of Eq.\,\eqref{eq:margden} requires a discretization of the continuous variable $X$ onto $N_X$ values. 
To do so, we use Eq.\,(6) from Ref.\,\cite{miao2013analysis}, for which the discretization of the Fokker-Planck equation reads 
\begin{align}
\hbar\partial_t
\left(
\begin{matrix*}[l]
P(X_{-1},\,t)\\ 
P(X_{0},\,t)\\  
P(X_{+1},\,t)
\end{matrix*}
\right)
= 
\hat{\Lambda}_{N_X=3} 
\left(
\begin{matrix*}[l]
P(X_{-1},\,t)\\ 
P(X_{0},\,t)\\  
P(X_{+1},\,t)
\end{matrix*}
\right)
= \hbar\lambda
\begin{pmatrix}
-1 & \frac{1}{2} & 0 \\ 
1 & -1 & 1 \\  
0 & \frac{1}{2} & -1 
\end{pmatrix}
\left(
\begin{matrix*}[l]
P(X_{-1},\,t)\\ 
P(X_{0},\,t)\\  
P(X_{+1},\,t)
\end{matrix*}
\right),
\end{align}
and the discretized stochastic field can take on the values 
\begin{align}
\left(
\begin{matrix*}[l]
X_{-1}\\ 
X_{0}\\  
X_{+1}
\end{matrix*}
\right)
=\sqrt{\kappa\lambda}
\left(
\begin{matrix*}[r]
-1\\ 
0\\  
1
\end{matrix*}
\right).
\end{align}
The associated steady state distribution is given by $P_\mathrm{ss} = (1/4,\,1/2,\,1/4)^T$.

For completeness we will state here explicitly the vectorization of Eq.\,\eqref{eq:margden} used for the main-text simulations.
For the column stacked vectorization we follow Ref.\,\cite{am2015three}: 
\begin{enumerate}
\item Coherent Dynamics
\begin{align}
-i[\hat{H},\hat{\rho}] \Leftrightarrow -i\left(\mathbbm{1}_{n} \otimes \hat{H} - \hat{H}^{T} \otimes \mathbbm{1}_{n}\right)\kett{\rho(t)} = -i\hat{\hat{H}}\kett{\rho(t)}
\end{align}
\item Dissipative Dynamics
\begin{align}
D[\hat{a}]\hat{\rho} \Leftrightarrow \left[(\hat{a}^{\dagger})^{T} \otimes \hat{a} - \frac{1}{2}\bigg(\left(\hat{a}^{\dagger}\hat{a}\right)^{T} \otimes \mathbbm{1}_n + \mathbbm{1}_n\otimes \hat{a}^{\dagger}\hat{a}\bigg)\right]\kett{\rho(t)} = \hat{\hat{D}}[\hat{a}]\kett{\rho(t)}
\end{align}
\end{enumerate}
Here, $\mathbbm{1}_{n}$ is the identity operator on the $n$-dimensional system Hilbert space.
Thus, the dynamics of the marginal densities is governed by 
\begin{align}
\hbar\partial_t 
\left(
\begin{matrix*}[l]
\kett{u(X_{-1},\,t)}\\
\kett{u(X_{0},\,t)}\\
\kett{u(X_{+1},\,t)}
\end{matrix*}
\right)
= \left[\hat{\Lambda}_3 \otimes \mathbbm{1}_{n^2} -i \hbar \sqrt{\kappa\lambda}
\begin{pmatrix}
-1 & 0 & 0\\
0 & 0 & 0\\
0 & 0 & +1\\
\end{pmatrix}
\otimes \hat{\hat{O}}
+
\mathbbm{1}_3\otimes
\left(
-i\hat{\hat{H}} + \hbar\sum_a \gamma_a \hat{\hat{D}}[\hat{a}] 
\right)
\right]
\left(
\begin{matrix*}[l]
\kett{u(X_{-1},\,t)}\\
\kett{u(X_{0},\,t)}\\
\kett{u(X_{+1},\,t)}
\end{matrix*}
\right).
\end{align}
The vectorized statistical average density operator is then calculated as $ \kett{\langle\rho(t)\rangle_\mathrm{s}} = \sum_\alpha \kett{u,\,X_\alpha}$. 
Note, beside the finite discretization of the continuous noise process this approach is exact. 

Finally we discuss the white noise limit of the OU process.
The white noise limit corresponds to the limit when the noise correlation time $1/\lambda$ vanishes ($\lambda \to \infty$) and thus the noise process becomes uncorrelated. 
In particular, this can be seen by evaluating the steady state noise covariance function 
\begin{align}
\langle  X(t) X(0) \rangle_\mathrm{s} -\langle X(t) \rangle_\mathrm{s}\langle X(0) \rangle_\mathrm{s} = \frac{\kappa\lambda}{2} e^{-\lambda t} \stackrel{\lambda\rightarrow \infty}{=} \kappa \delta (t).
\end{align}
Note, this analysis applies to the continuous and discretized OU process. 
The strength of the white noise process is determined by $\kappa$ and the associated master equation reads
\begin{equation}
\label{eq_NoisyMasterEquation}
\hbar\partial_t\hat{\rho}(t) =-i [\hat{H}, \hat{\rho}(t)] 
+ \hbar\sum_{a} \gamma_a \mathcal{D}\left[\hat{a}\right]\hat{\rho}(t) + \hbar\kappa \mathcal{D}\big[ \hat{O}\big]\hat{\rho}(t),
\end{equation}
which is used for the white noise simulation presented in the main-text.

\section{Pauli transfer matrix}
\label{C}

In the following section we discuss time-evolution of logical states using the Pauli transfer matrix (PTM) formalism. 
This formalism is used in the main-text to characterize the error channels, when the system is subject to colored noise processes and imperfect logical gate operations (described in subsequent sections).  

A logical initial state, given by a density matrix $\hat{\rho}(t=0)$, can be expressed as
\begin{align}
\hat{\rho}(0) = \bigg[d_0(0)\,\widetilde{\mathbb{1}} +  \!\!\!\!\!\!\sum_{n\in\{\widetilde{x}, \widetilde{y}, \widetilde{z}\}} \!\!\!\!\! d_n(0) \hat{\sigma}_{n}\bigg]/2,
\end{align}
where $\widetilde{\mathbb{1}}$ and $\hat{\sigma}_{\widetilde{x}}$, $\hat{\sigma}_{\widetilde{y}}$, and $\hat{\sigma}_{\widetilde{z}}$ are the logical identity- and Pauli-operators, respectively, as defined in the main-text. 
Furthermore, $\boldsymbol{d}(0) = (d_{\widetilde{x}}(0),\, d_{\widetilde{y}}(0),\,d_{\widetilde{z}}(0) )^\mathrm{T}$ is the three component logical Bloch sphere vector, with $|\boldsymbol{d}(0)|\leq 1$.
For completeness we also keep the fourth component $d_0(0)$, which describes leakage out of the logical subspace when $d_0(0)<1$.
Time evolution from $\hat{\rho}(0)$ to $\hat{\rho}(t)$ can be expressed as \cite{roncallo2023pauli}
\begin{align}
\hat{\rho}(t) &= \bigg[d_0(t)\,\widetilde{\mathbb{1}} +  \!\!\!\!\!\!\sum_{n\in\{\widetilde{x}, \widetilde{y}, \widetilde{z}\}} \!\!\!\!\! d_n(t) \hat{\sigma}_{n}\bigg]/2,\text{ with }\\
d_n(t) &= \sum_m R_{n,m}(t) d_m(0),
\end{align}
where $R_{n,m}(t)$ is the PTM given by
\begin{equation}
R_{n,m}(t) = \frac{1}{2}\text{tr}\left[ \hat{E}_n\, \hat{\rho}_m(t) \right] \text{ for } \hat{\rho}_m(0) =\hat{E}_{m},
\end{equation}
and the operators $\hat{E}_{n}$ are defined as 
\begin{equation}
\begin{aligned}
\hat{E}_{0} &= \widetilde{\mathbb{1}} = \ket{\widetilde{0}}\bra{\widetilde{0}} + \ket{\widetilde{1}}\bra{\widetilde{1}},\\
\hat{E}_{\widetilde{x}}&= \hat{\sigma}_{\widetilde{x}} = \ket{\widetilde{1}}\bra{\widetilde{0}} + \ket{\widetilde{0}}\bra{\widetilde{1}},\\
\hat{E}_{\widetilde{y}}&= \hat{\sigma}_{\widetilde{y}} = -i\left(\ket{\widetilde{1}}\bra{\widetilde{0}} - \ket{\widetilde{0}}\bra{\widetilde{1}}\right),\text{ and }\\
\hat{E}_{\widetilde{z}}&= \hat{\sigma}_{\widetilde{z}} = \ket{\widetilde{1}}\bra{\widetilde{1}} - \ket{\widetilde{0}}\bra{\widetilde{0}}.
\end{aligned}
\end{equation}

The diagonal elements $R_{n,n}(t)$ decrease at a rate dependent on the noise strength and correlation time. A decrease of the diagonal elements of the PTM characterizes shrinkage of the Bloch vector, i.e. increases the mixedness of quantum state.

\section{Autonomous stabilization}
\label{sec_AutonomousStabilization}


Here we detail the general theory employed for our autonomous stabilization protocol, that protect the logical qubit from the colored noise models discuss section \ref{sec:Colored}. 
We consider the laser coupling Hamiltonian from Eq.\,\eqref{SM:H} with zero detunings and Rabi frequencies $\Omega^{+1} = -\Omega^{-1}=-\sqrt{2}\Omega$,
\begin{equation}
\hat{H}_{\mathrm{DS}}/\hbar = -\frac{\Omega}{\sqrt{2}} \left(\hat{\mathcal{C}}_{+1} - \hat{\mathcal{C}}_{-1} + \mathrm{h.c.}\right),
\end{equation}
for which the two DSs $\hat{H}_{\mathrm{DS}}\ket{\mathrm{DS}_{1,2}} = 0$ [corresponding to our logical qubit states from the main-text Eq.\,(2)] are,
\begin{equation}
\begin{aligned}
\label{eq:dark-states}
\ket{\mathrm{DS}_1} &= \ket{\widetilde{0}} = e^{-i \frac{\pi}{2} \hat{F}_{g,y}} \ket{F_g, +F_g} = \ket{F_g, +F_g}_x~\mathrm{and}\\
\ket{\mathrm{DS}_2} &=\ket{\widetilde{1}} = e^{-i \frac{\pi}{2} \hat{F}_{g,y}} \ket{F_g, -F_g} =\ket{F_g, -F_g}_x
\end{aligned}
\end{equation}
according to Sec.\,\ref{sec:DS}.
Here, $\ket{F_g, m}_{x}$ is a hyperfine state with angular momentum projection $m$ along the $x$-axis of the Bloch sphere (rather than along the magnetic quantization axis). 
In addition to laser driving the $F_e$-manifold is subject to spontaneous emission \cite{sobel2016introduction}, described by the following master equation 
\begin{align}
\label{eq_StabMasterEquation}
&\hbar\partial_t\hat{\rho}(t) = \hat{\mathcal{L}}_{\mathrm{stab}}\hat{\rho}(t) = -i [\hat{H}_{\mathrm{DS}}, \hat{\rho}(t)] 
+ \hbar\gamma \!\! \sum_{q=0,\pm 1} \mathcal{D}\left[\hat{\mathcal{C}}_q \right]\hat{\rho}(t).
\end{align}
As spontaneous emission only affects the $F_e$-manifold of states, the four operators $|\widetilde{0}\rangle\langle\widetilde{0}|$, $|\widetilde{1}\rangle\langle\widetilde{0}|$, $|\widetilde{0}\rangle\langle\widetilde{1}|$ and $|\widetilde{1}\rangle\langle\widetilde{1}|$ are the steady states of the system.
Thus, in the limit of infinite time-evolution $t \to \infty$ under the above driven-dissipative system, any state $\hat{\rho}$ will relax to a density matrix spanning just the logical subspace, $\hat{\rho}_\infty \equiv \lim_{t \rightarrow \infty} \hat{\rho}(t)$ given by,
\begin{equation}
\label{eq_projectIntoLogicalSubspace}
\hat{\rho}_\infty = c_{00} \ket{\widetilde{0}}\bra{\widetilde{0}} + c_{01} \ket{\widetilde{0}}\bra{\widetilde{1}} + c_{10} \ket{\widetilde{1}}\bra{\widetilde{0}} + c_{11}\ket{\widetilde{1}}\bra{\widetilde{1}}.
\end{equation}
The coefficients $c_{\mu \nu}$ are determined by the overlap of the initial density matrix with the system's appropriate conserved quantities, 
\begin{equation}
\label{eq_conservedQuantityTrace}
c_{\mu\nu} = \text{tr}\left[\hat{J}_{\mu\nu}^{\dagger} \hat{\rho}(0)\right],
\end{equation}
where $\hat{J}_{\mu\nu}$ are the left hand eigenvectors of the Liouvillian with eigenvalue zero
\begin{equation}
\bbra{J_{\mu\nu}} \hat{\hat{\mathcal{L}}}_{\mathrm{stab}} = 0,
\end{equation}
where $\kett{A}$ is a vectorized (columns stacked) version of an operator $\hat{A}$, and the corresponding Liouvillian super-operator matrix is defined via the master equation $\partial_t \kett{\rho} =\hat{\hat{\mathcal{L}}}_{\mathrm{stab}}\kett{\rho}$~\cite{albert2018lindbladians}.

The conserved quantities associated with the $\hat{\mathcal{L}}_{\mathrm{stab}}$ can be found analytically ($\delta=\Delta=0$) in both integer and half-integer spin $F_g$,
\begin{equation}
\label{eq:conserved-quantity}
\begin{aligned}
\hat{J}_{00} &= \ket{\widetilde{0}}\bra{\widetilde{0}} + \sum^{F_g-1}_{m=-F_g +1} a_m \bigg(\ket{F_g,m}_x\!\bra{F_g,m} + \ket{F_e,m}_x \!\bra{F_e,m} \bigg), \\
\hat{J}_{11} &= \mathbbm{1}- \hat{J}_{00},\\
\hat{J}_{01} &= \ket{\widetilde{0}} \bra{\widetilde{1}},\\
\hat{J}_{10} &= \ket{\widetilde{1}} \bra{\widetilde{0}},
\end{aligned}
\end{equation}
with $\mathbbm{1} = \sum_{m=-F_g}^{F_g}\ket{F_g,m}_x\! \bra{F_g,m} + \sum_{m=-F_e}^{F_e} \ket{F_e,m}_x\! \bra{F_e,m}$ is the identity operator. The coefficients $a_m$ for $1-F_g \leq m \leq F_g-1$ satisfies
\begin{equation}
\label{eq: am-J00}
\begin{aligned}
a_m &= \frac{ \sum_{m'=-F_g}^{m-1} G(m') }{ \sum_{m'=-F_g}^{F_g-1} G(m') },\,\mathrm{with}\\
G(m) &=
\begin{cases}
    \prod_{m'=-F_g+1}^{m} \frac{(F_g-m'+1)(F_g-m')}{(F_g+m'+1)(F_g+m')}, \quad & -F_g+1 \leq m \leq F_g-1, \\
    1, \quad & m = -F_g.
\end{cases}
\end{aligned}
\end{equation}
The denominator in the expression of $a_m$ can be further evaluated as
\begin{equation}
    \sum_{m'=-F_g}^{F_g-1} G(m') = \Hypergeometric{2}{1}{-2F_g, 1-2F_g}{2}{1} = \frac{(4F_g)!}{(2F_g+1)! (2F_g)!}\simeq\frac{16^{F_g}}{F_g^{3/2}\sqrt{8\pi}} \qquad (F_g \gg 1),
\end{equation}
where $\Hypergeometric{2}{1}{a,b}{c}{z}$ is the hypergeometric function, and the approximated expression in the end is achieved using Stirling's formula $n! \simeq \sqrt{2\pi n}(\frac{n}{e})^n$ under the large $n$ limit.

With the knowledge of the conserved quantities, we can evaluate the action of the autonomous stabilization protocol following various logical operations without the need to explicitly evolve the master equation to long times. 
Whenever we employ stabilization (and no other non-negligible Hamiltonians or dissipators), we evaluate the effect by simply projecting the density matrix of the qubit into the logical subspace via Eqs.~\eqref{eq_projectIntoLogicalSubspace} and \eqref{eq_conservedQuantityTrace}.

\subsection{Biased effective error model}

While we use a specific colored noise simulation to benchmark the 'dark spin-cat' against unwanted external noise, the exponential suppression of bit-flip errors in the cat size $F_g$ emerges quite generically for \textit{any} perturbation that acts locally in the basis of spin states, i.e. that changes angular momentum by only a few units at a time. 
As an example, we consider white noise generated by $\mathcal{D}[\sqrt{\kappa}\hat{F}_{g,z}]$. 
The conserved quantities $\hat{J}_{nm}$ can be used to derive the exponentially in $F_g$ suppressed change of rate of the PTM, which is to leading first order in $\kappa$ given by 
\begin{align} 
    \dot{R}_{\widetilde{z}, \widetilde{z}}  = -\kappa F_g a_{-(F_g-1)}  \simeq - \kappa \frac{F_g^{5/2}\sqrt{8\pi}}{16^{F_g}}.
\end{align}
On the contrary, the logical dephasing error rate only amplifies by a factor of $F_g$. The effect of $\mathcal{D}[\sqrt{\kappa}\hat{F}_{g,y}]$ will be similar to $\mathcal{D}[\sqrt{\kappa}\hat{F}_{g,z}]$ due to the rotational symmetry along $x$ direction. On the other hand, $\mathcal{D}[\sqrt{\kappa}\hat{F}_{g,x}]$ noise will not let the code states leave out of the encoded subspace but will give $\ket{\widetilde{0}}, \ket{\widetilde{1}}$ a relative phase, which leads to a dephasing error with the rate scales as $O(F_g^2)$.

\begin{figure}[h]
    \centering
    \includegraphics[width=0.8\linewidth]{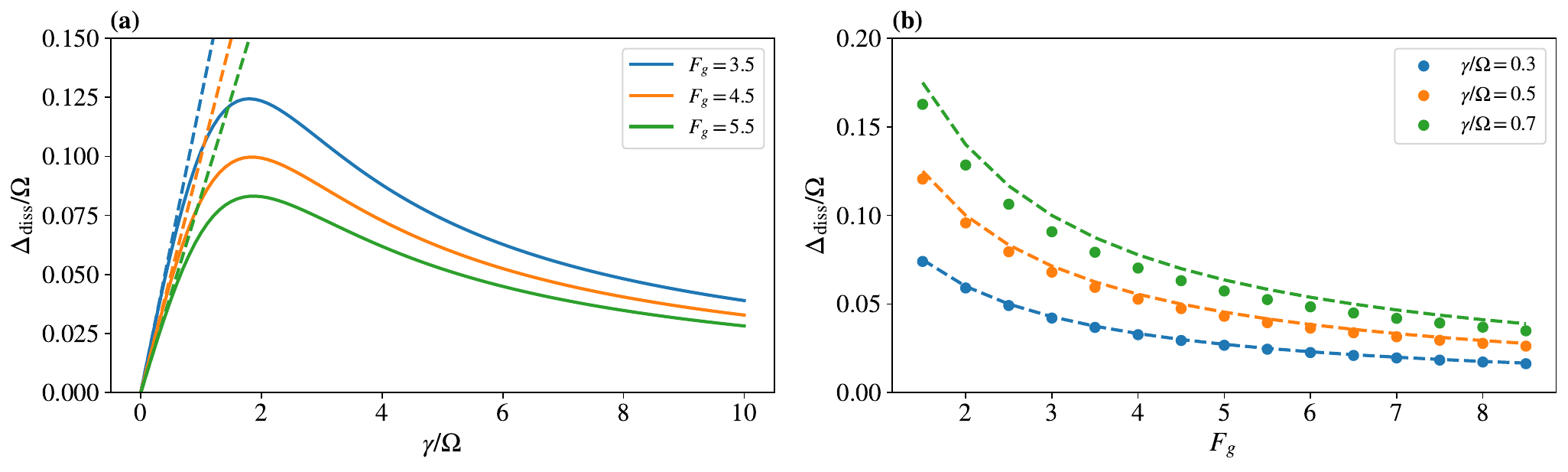}
    \caption{Dissipation gap $\Delta_\text{diss}$ varies as the change of spontaneous emission strength $\gamma$ and spin length $F_g$. (a) Fixed $F_g$ and varying $\gamma$. (b) Fixed $\gamma$ and varying $F_g$. Dashed lines in both plots are references with $\Delta_\text{diss} = \gamma/(2F_g+1)$.}
    \label{fig:diss_gap}
\end{figure}

As a practical aside, while the autonomous stabilization will always relax the qubit into the logical subspace in the limit $t \to \infty$, the rate of stabilization is set by the dissipative gap $\Delta_{\mathrm{diss}}$. 
This gap can be written as $\Delta_{\mathrm{diss}} = - \text{Re}(\lambda_{\mathrm{diss}})$, where $\lambda_{\mathrm{diss}}$ is the right eigenvalue of the Liouvillian super-operator  $\hat{\hat{\mathcal{L}}}_{\mathrm{stab}} \kett{\lambda_{\mathrm{diss}}} =\lambda_{\mathrm{diss}}\kett{\lambda_{\mathrm{diss}}}$, with the smallest non-zero real part. 
A larger $\Delta_{\mathrm{diss}}$ leads to faster stabilization. If the stabilization is active at the same time as an error process with magnitude $\kappa$ is present, as considered in the main-text, the error will be suppressed provided $\kappa F_g \ll \Delta_{\mathrm{diss}}$.
In our simulations, we consider the regime $|\mathbf{\Omega}| \gg \gamma$, where to numerical study suggests that dissipation gap satisfies $\Delta_{\operatorname{diss}} \simeq \gamma/(2F_g+1)$, as shown in FIG.~\ref{fig:diss_gap}.

\subsection{Numerical benchmark}
Here we comment on the numerical simulation of the autonomous stabilization mechanism presented in the main-text. 
In particular, in the main-text Figs.~1(c) and~3(b), we present the time derivative of the diagonal elements of the PTM $\partial_t R_{n,n}(t)$ rather than the matrix elements themselves when subjected to a colored-noise process as discussed in Sec.\,\ref{sec:Colored}.
Note, the element $R_{0,0}(t)$ describes leakage out of the logical qubit subspace and takes on for $\kappa t\rightarrow \infty$ a constant value close to unity, \emph{i.e.} a balance between the leakage generating colored noise process and the autonomous stabilization mechanism.

\begin{figure}[!t]
\center
    \includegraphics[width=0.6\columnwidth]{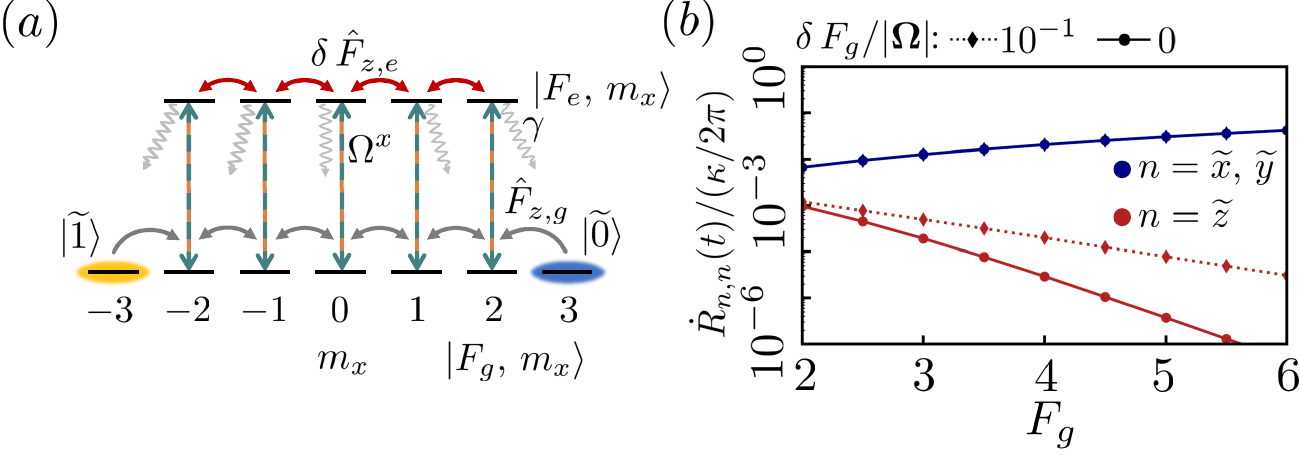}
\caption{(a) Illustration of laser coupling (\mbox{$F_g\!=\!3$}) in the rotated coordinate system, for which \mbox{$\boldsymbol{\Omega}=(\Omega^x,\,0,\,0)^T$} defines the quantization axis. Note, the detunings are taken in contrast to the main-text figure 3(a) to be $\Delta\!=\! 0$ and $\delta\neq 0$, which gives rise to an additional coupling in the $F_e$-manifold of states (red arrows).
Spontaneous emission of $|F_{e},\,m_x\rangle$ is due to the CG coefficients directed "towards" the closest DS, as illustrated by wavy arrows indicating the average $m_x$ change.
(b) Time derivative of the diagonal elements of $\mathbf{R}(t)$ as a function of $F_g$ for different values of $\delta$.
Here $\Delta\!=0$, $|\boldsymbol{\Omega}|/\gamma\!=\!2\pi$, $F_g\!=\!4$, $\kappa\!=\!10^{-4}|\boldsymbol{\Omega}|$, $\lambda/|\boldsymbol{\Omega}|=10^{-3}$ and $N_X\!=\!3$.}
\label{fig:SM2}
\end{figure}

For the main-text simulations we assumed the Land\'e-g factor of the $F_g$ and $F_e$ manifold of states are equal, thus we use $\delta=0$ in the main-text Eq.\,(1).
The actual value of $\delta$ depends on the specific atomic platform, the considered levels and also the applied magnetic field. 
In the following we will discuss the effect of such a non-zero $\delta$.
For the parameters chosen in the main-text below Eq.\,(2) ($\boldsymbol{\Omega}=(\Omega^x,\,0,\,0)^T$) the laser-coupling becomes particularly simple, see Fig.\,\ref{fig:SM2} (or main-text Fig.\,3(a)). 
In this frame the differential shift $\delta\,F_{e,z}$ couples adjacent states in the $F_e$-manifold of states. 
This coupling alters the composition of the bright-states and, thus, disturbs the autonomous stabilization mechanism. 
In particular, the exponent of the exponentially suppressed bit-flip error rate $R_{\widetilde{z},\widetilde{z}}(t)$ is increased, whereas, the dephasing rates $R_{\widetilde{x},\widetilde{x}}(t)$ and $R_{\widetilde{y},\widetilde{y}}(t)$ are unaltered,
see Fig.\,\ref{fig:SM2} (b). 
However, from numerical simulations we observe that there is still an exponential suppressed bit-flip error rate as compared to the dephasing-rate for $\delta\,F_g<|\boldsymbol{\Omega}|$, hence, the model-system still exhibits an exponentially in $F_g$ biased error model. 

\subsection{Autonomous error correction with engineered dissipation}

Above we discussed the ability of autonomous stabilization in the system we considered, i.e., when the population leaves the encoded subspace due to the perturbation from the noise, it will passively go back under the dissipative process. However, as indicated in the form of conserved quantities $\hat J_{01}$ and $\hat J_{10}$ shown in Eq.~\eqref{eq:conserved-quantity}, if initially there is quantum coherence between $\ket{\widetilde{0}}$ and $\ket{\widetilde{1}}$, then such coherence will be totally lost when it leaves the code subspace and goes back again due to the stabilization. Therefore, the leakage will be converted to a dephasing error.

\begin{figure}[t]
    \centering
    \includegraphics[width=0.3\linewidth]{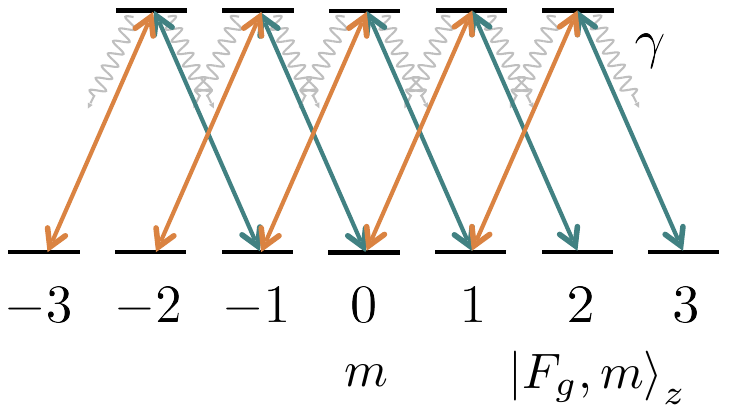}
    \caption{Schematic plot for the engineered dissipation. When atoms are in a cavity, the decay may happen with two different polarizations rather than three. The system can be separated into two parts characterized by parity $\hat \Pi$, where there is no coupling between the two parts.}
    \label{fig:engi-dissi}
\end{figure}

On the other hand, if we can select the polarization of the decay from excited levels to the ground, then certain noise like $\mathcal{D}[\sqrt{\kappa}\hat{F}_{g,z}]$ can be further corrected. The idea is to only keep the decay process with $\sigma_\pm$ polarization ($\Delta m_z = \pm 1$, see Fig.~\ref{fig:engi-dissi}), which can be achieved by putting the atoms in a cavity~\cite{pineiro_orioli_emergent_2022}. In this way, the dissipative part in the master equation \eqref{eq_StabMasterEquation} should be modified to $\gamma \sum_{q=\pm 1} \mathcal{D}[\hat{\mathcal{C}}_q]\hat{\rho}(t)$. Now the parity operator $\hat \Pi = e^{i\pi (\hat F_{g,z} + \hat F_{e,z} +\hat P_e - F_g) }$ will be preserved during the evolution, so for noise like $\mathcal{D}[\sqrt{\kappa}\hat{F}_{g,z}]$ that does not affect the parity of the states, the coherence between $\ket{\widetilde{0}}$ and $\ket{\widetilde{1}}$ can also be mostly preserved, in contrast to the case where the excited manifold is subject to spontaneous decay. However, the dephasing error induced by $\mathcal{D}[\sqrt{\kappa}\hat{F}_{g,y}]$ will increase, as it will flip the parity of the code states while causing them to leave the encoded subspace.

Those features can also be illustrated with the analysis of conserved quantities under the new Lindbladian. In the strong driving limit ($\Omega \gg \gamma$), we can still write down the expressions for those conserved quantities to the 0-th order (up to $O[(\gamma/\Omega)^0]$) as the following:
\begin{equation}
\label{eq:conserved-quantity}
\begin{aligned}
\hat{J}'_{00} &\simeq \ket{\widetilde{0}}\bra{\widetilde{0}} + \sum^{F_g-1}_{m=-F_g +1} a_m \bigg(\ket{F_g,m}_x\!\bra{F_g,m} + \ket{F_e,m}_x \!\bra{F_e,m} \bigg), \\
\hat{J}'_{11} &= \mathbbm{1}- \hat{J}'_{00},\\
\hat{J}'_{01} &\simeq \ket{\widetilde{0}} \bra{\widetilde{1}}+ \sum^{F_g-1}_{m=-F_g +1} a_m \bigg(\ket{F_g,m}_x\!\bra{F_g,-m} + \ket{F_e,m}_x \!\bra{F_e,-m} \bigg),\\
\hat{J}'_{10} &= \hat J_{01}^{\prime\dagger}.
\end{aligned}
\end{equation}
For a state $\ket{\psi} = a\ket{\widetilde{0}} + b\ket{\widetilde{1}}$, it will hop to $\ket{\psi_1} = a\ket{F_g, F_g-1}_x + b\ket{F_g, -(F_g-1)}_x$ when suffering from a jump operator $\hat F_{g,z}$. However, the difference between the overlap $\bra{\psi}\hat{J}'_{01}\ket{\psi}$ and $\bra{\psi_1}\hat{J}'_{01}\ket{\psi_1}$ is again exponentially suppressed with $F_g$, which indicates that the phase coherence between $\ket{\widetilde{0}}$ and $\ket{\widetilde{1}}$ can also be well preserved after the dissipative stabilization process. This observation, together with the bit-flip protection, provides a passive correction of the $\hat F_{g,z}$ error with the engineered dissipation.

\section{Numerical simulations of logical gates}
\label{E}

Here we detail the numerical simulations used to benchmark the adiabatic logical gates in the main-text Figure 3. 
The same colored-noise simulation described for $N_X=3$ from above is employed, except the driving laser Rabi frequency phases and amplitudes are explicitly time-dependent.
Therefore, we consider the generalised time-dependent Hamiltonian from Eq.\,\eqref{SM:H}, namely
\begin{equation}
\hat{H}_{\mathrm{DS}}(t)/\hbar = \frac{1}{2}\sum_{q = 0, \pm 1}\left(\Omega^{q}(t)\hat{\mathcal{C}}_q+\mathrm{h.c.}\right)
\label{eq:Hamiltonian_gate}
\end{equation}
From what will follow, it is convenient to parameterize the Rabi frequencies as follows
\begin{align}
\label{eq:driveParametrization}
\begin{pmatrix}
\Omega^{+1}\\
\Omega^{0}\\
\Omega^{-1}
\end{pmatrix}
=\Omega
\begin{pmatrix}
 -\frac{1}{\sqrt{2}}e^{i \alpha(t)}\! &\sin \left[\beta(t)\right]  \\
 &\cos \left[\beta(t)\right] \\
 \frac{1}{\sqrt{2}}e^{-i \alpha(t)}\! &\sin \left[\beta(t)\right]
\end{pmatrix},
\end{align}
with time dependent parameters $\alpha(t)$ and $\beta(t)$.
The associated Cartesian components of the Rabi frequencies 
\begin{align}
\Omega^x &= \Omega\cos[\alpha(t)]\sin[\beta(t)],\notag\\ 
\Omega^y &= \Omega\sin[\alpha(t)]\sin[\beta(t)],~\mathrm{and}\notag\\
\Omega^z &= \Omega\cos[\beta(t)]
\end{align}
are by construction always real and, thus, the  instantaneous DSs $|\mathrm{DS}_{1,2}(t)\rangle$, satisfying $\hat{H}_{\mathrm{DS}}(t) |\mathrm{DS}_{1,2}(t)\rangle=0$, can always be identified with orthogonal SCSs, see Sec.\,\ref{sec:DS}
\begin{align}
\ket{\mathrm{DS}_1(t)} &= \hat{\mathcal{R}}[\alpha(t),\,\beta(t),0]\ket{F_g, +F_g},~\mathrm{and}\notag\\
\ket{\mathrm{DS}_2(t)} &= \hat{\mathcal{R}}[\alpha(t),\,\beta(t),0]\ket{F_g, -F_g}.
\label{DS1,2}
\end{align}
The orthogonality of the instantaneous DS gives rise to exponentially in $F_g$ supressed bit-flip error rates. 
This is because any fixed configuration of laser couplings with antipodal dark states is inherently bias preserving as per our analysis in the main text. 
A global SU(2) rotation of the frame does not change this property, provided the variation of the underlying laser couplings is sufficiently adiabatic.
Note, after after simulating each gate operation via the colored noise master equation, we perform autonomous stabilization.
To avoid excessive computation, we instead project the system into the subspace of driven-dissipative DSs directly as per Section~\ref{sec_AutonomousStabilization}.

\subsection{$\hat{U}_{\widetilde{z}}(\alpha_z)$ gate}

We start with the single-qubit rotation $\hat{U}_{\widetilde{z}}(\alpha_z) = \exp (-i \hat{\sigma}_{\widetilde{z}}\alpha_z/2)$ for a fixed rotation angle $\alpha_z$ [not to be confused with the time-dependent laser parameter $\alpha(t)$]. The methodology for the gate is to adiabatically vary the direction of the DSs on the Bloch sphere [parametrized by the angles $\alpha(t)$, $\beta(t)$ from Eq.~\eqref{eq:driveParametrization}] along a closed loop, as depicted in Main-text Fig. 2(b). The solid angle on the Bloch sphere enclosed by the loop sets the accrued phase and hence rotation angle $\alpha$.

The specific laser ramp profile we choose, as a function of time $t \in [0,T]$ with $T$ the total gate time, is given by,
\begin{equation}
\begin{aligned}
\begin{array}{ll}
 0 < t < \frac{T}{8}: & \begin{array}{l}
\alpha(t) = 0 \\
\beta(t) = \frac{\pi}{2} - \beta_1\left[1 + \sin\left(4\pi\frac{t}{T} - \frac{\pi}{2}\right)\right]
\end{array} \\
\\
\frac{T}{8} < t < \frac{T}{2}: & \begin{array}{l}
\alpha(t) = \alpha_1 \cdot \frac{8}{3}(\frac{t}{T} - \frac{1}{8}) \\
\beta(t) = \frac{\pi}{2} - \beta_1
\end{array} \\
\\
\frac{T}{2} < t < \frac{5T}{8}: & \begin{array}{l}
\alpha(t) = \alpha_1 \\
\beta(t) = \frac{\pi}{2} - \beta_1\left[1 + \sin\left(4\pi\frac{t}{T} - \pi\right)\right]
\end{array} \\
\\
\frac{5T}{8} < t < T: & \begin{array}{l}
\alpha(t) = \alpha_1 \cdot -\frac{8}{3}(\frac{t}{T} - 1) \\
\beta(t) = \frac{\pi}{2}
\end{array}
\end{array}
\end{aligned}
\end{equation}
where $\alpha_1 = \frac{5\pi}{26}, \beta_1 = \frac{5\pi}{78}$ are fixed parameters. This profile yields a rotation angle $\alpha_z = 2 F_g \alpha_1 \cos(\frac{\pi}{2}-\beta_1)$. The profile is shown in main- text Fig.2(c).

Note that the PTM $\mathbf{R}$ for this gate operation is not diagonal. Ideally, it takes the form of,
\begin{equation}
\mathbf{R}_{\mathrm{ideal}} = \left(\begin{array}{cccc}
1 & 0 & 0 & 0 \\
0 & \cos(\alpha_z) & \sin(\alpha_z) & 0 \\
0 & \sin(\alpha_z) & -\cos(\alpha_z) & 0 \\
0 & 0 & 0 & 1
\end{array}\right).
\end{equation}
In our simulations, we compute the ``raw" PTM via the colored-noise simulations, multiply it by this ideal matrix $\mathbf{R}\to \mathbf{R} \mathbf{R}_{\mathrm{ideal}}^{-1}$, and minimize the off-diagonal elements over all $\alpha_z$ (which amounts to correcting for deterministic under- or over-rotation, and can be corrected by adjusting the overall angle $\alpha_z$ of the gate). The remaining diagonal elements characterize the unrecoverable error channel of the gate, and are reported in main- text Fig.\,3(c).

One can also add a counter-diabatic Hamiltonian
$\hat{H}_{\text{CD}}(t)$ to compensates for the rapid driving by effectively cancelling the non-adiabatic transitions \cite{del2013shortcuts}.
\begin{equation}
    \hat{H}_{\text{CD}}(t)/\hbar = i\sum_{n = 1,2} \left( \ket{\partial_t \mathrm{DS}_{n}(t)} \bra{\mathrm{DS}_{n}(t)} - \bra{\mathrm{DS}_{n}(t)}  \partial_t \mathrm{DS}_{n}(t)\rangle \ket{\mathrm{DS}_{n}(t)} \bra{\mathrm{DS}_{n}(t)} \right)
\end{equation}
where $\ket{\mathrm{DS}_{n}(t)}$ are the instantaneous eigenstates of the Hamiltonian $\hat{H}_{\mathrm{DS}}(t)$, defined in Eq.\,\eqref{DS1,2}. For the $\hat{U}_{\widetilde{z}}(\alpha_z)$ gate, we can add a linear combination of $\hat{F}_{g,x}, \hat{F}_{g,y}, \hat{F}_{g,z}$ drive to actively rotate the $\ket{\widetilde{0}}$, $\ket{\widetilde{1}}$ logical states to make sure it follows the adiabatic loop. 

\begin{equation}
\begin{aligned}
\begin{array}{ll}
 0 < t < \frac{T}{8}: & \begin{array}{l}
\hat{H}_{\text{CD}}(t) = \dot{\beta}(t) \hat{F}_{g,y}
\end{array} \\
\\
\frac{T}{8} < t < \frac{T}{2}: & \begin{array}{l}
\hat{H}_{\text{CD}}(t) = \cos(\alpha(t))\sin(\beta_1)\cos(\beta_1)\dot{\alpha}(t) \hat{F}_{g,x} \\
\quad + \sin(\alpha(t))\sin(\beta_1)\cos(\beta_1)\dot{\alpha}(t) \hat{F}_{g,y} + \sin(\alpha(t))\cos(\beta_1)\dot{\alpha}(t) \hat{F}_{g,z}
\end{array} \\
\\
\frac{T}{2} < t < \frac{5T}{8}: & \begin{array}{l}
\hat{H}_{\text{CD}}(t) = \dot{\beta}(t) \hat{F}_{g,x}
\end{array} \\
\\
\frac{5T}{8} < t < T: & \begin{array}{l}
\hat{H}_{\text{CD}}(t) = \dot{\alpha}(t) \hat{F}_{g,z}
\end{array}
\end{array}
\end{aligned}
\end{equation}
This profile transforms the approximate adiabatic solution into the exact solution of time dependent Schr\"odinger equation. 

Note, the instantaneous counter-diabatic terms, which cancel the diabatic terms, are always orthogonal to the instantaneous axis defining the SCS. Thus, there exists a spatial rotation (around the SCS encoding axis) for which diabatic corrections act like an additional magnetic field.

\subsection{$\hat{U}_{\widetilde{x}}$ gate}

Next we consider a $\hat{U}_{\widetilde{x}} = \exp (- i \pi \hat{\sigma}_{\widetilde{x}})$ operation. This is straightforward to accomplish, as it amounts to rotating the state by $\pi$ along the equator of the Bloch sphere, or equivalently swapping the $\ket{\widetilde{0}}$, $\ket{\widetilde{1}}$ logical states. One can accomplish the former by simply time-varying the laser couplings as,
\begin{align}
0 < t < T: ~~\alpha(t) = \pi\,t/T,~\mathrm{and}~\beta(t) = \pi/2
\end{align}
which corresponds to $\Omega^{x} (t) = \Omega \cos(\pi t/T)/\sqrt{2}$, $\Omega^{y}(t) = \Omega \sin(\pi t /T)/\sqrt{2}$ as shown in the main- text Fig.\,2(c). The fidelity performance of this gate is similar to that of the $\hat{U}_{\widetilde{z}}(\alpha_z)$ gate, hence we do not provide explicit benchmarks. Furthermore, this gate can also be implemented digitally by an exchange of the $\{\ket{\widetilde{0}},\ket{\widetilde{1}}\}$ logical states for the corresponding qubit.
Doing so avoids the need for explicit (potentially faulty) quantum operations.

Similar to the $\hat{U}_{\widetilde{z}}$ gate, one can add a counter-adiabatic Hamiltonian
\begin{align}
0 < t < T: ~~\hat{H}_{\text{CD}} = \dot{\alpha}(t) \hat{F}_{g,z}
\end{align}
 to speed up the gate.

\subsection{State preparation}

Next we describe the preparation of a qubit into a logical state $\ket{\widetilde{+}} = (\ket{\widetilde{0}} + \ket{\widetilde{1}})/\sqrt{2}$. The protocol starts by preparing a state in a maximally polarized state of the qubit hyperfine manifold $\ket{\psi_g} = \ket{F_g, -F_g}$ along the magnetic quantization axis. Atoms can be loaded into such a state with standard optical-pumping techniques.

\begin{figure}
\centering
\includegraphics[width=0.45\textwidth]{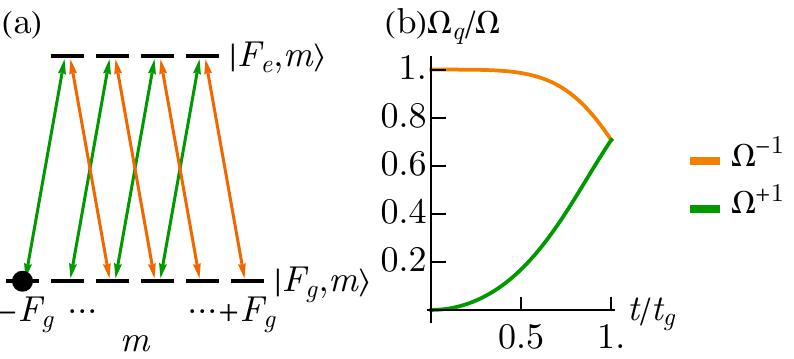}
\caption{(a) Schematic of the lasers and states used for preparation of the logical $\ket{\widetilde{+}} = (\ket{\widetilde{0}} + \ket{\widetilde{1}})/\sqrt{2}$ state, starting from a spin-polarized $\ket{F_g,-F_g}$ atom. (b) Laser ramp profile for state preparation.}
\label{fig:Preparation}
\end{figure}

We then turn the circular-polarized lasers $\Omega^{-1}(t)$ and $\Omega^{+1}(t)$ on, as depicted in Fig.~\ref{fig:Preparation}(a). The time-dependent profile of the lasers is chosen analogous to a STIRAP protocol, where the laser that does not couple to the initially-populated state $\ket{F_g,-F_g}$ is turned on first. Fig.~\ref{fig:Preparation}(b) shows the time-dependence of the laser couplings,
\begin{equation} \label{eq:OmegaStatePrep}
\begin{aligned}
\Omega^{-1}(t) & = \Omega \frac{1 + \cos\theta(t)}{\sqrt{2(1+\cos^2\theta(t))}}, \\
\Omega^{+1}(t) & = -\Omega \frac{1 - \cos\theta(t)}{\sqrt{2(1+\cos^2\theta(t))}},
\end{aligned}
\end{equation}
where in our simulations $\theta(t):=\frac{\pi}{2}\frac{t}{T}$ with $t = 0$ to preparation time $T$.
During this adiabatic process, the two non-orthogonal SCSs stabilized by the system are
\begin{equation}
\ket{\mr{SCS}_1} = \ket{\pi-\theta(t),0}, \quad \ket{\mr{SCS}_2} = \ket{\pi-\theta(t),\pi}.
\end{equation}
So at time $T$ the state we prepare will be a superposition of $\ket{\widetilde{0}}$ and $\ket{\widetilde{1}}$. Further, from the parity consideration the final state will be $\ket{\widetilde{+}} = (\ket{\widetilde{0}} + \ket{\widetilde{1}})/\sqrt{2}$ given sufficient adiabaticity of the evolution.

Note that for this operation we directly report the fidelity rather than the PTM in the main-text Fig.3(d), as such matrix elements can exhibit ambiguity depending on the ramping state preparation protocol, since the operation extends outside the logical subspace. The master equation is otherwise solved in the same way as before, also employing colored noise with $N_X = 3$.

\subsection{$\hat{\mathrm{C}}_X$ entangling gate}

Here we detail the numerical simulation used to benchmark the $\hat{\mathrm{C}}_X$ gate implemented via the Rydberg blockade mechanism. 
As described in the main-text, the $\hat{\mathrm{C}}_X$ gate operation involves two atoms labeled control ($\mathrm{C}$) and target ($\mathrm{T}$).
The control and target atom's logical qubit states are considered to be encoded x-axis within the $F_g$ manifold of states, see main-text equation (2). 
The $F_g$ manifold of state is laser-coupled to a Rydberg manifold of states $\ket{F_r,m}_{\mathrm{T}}$, with $m\in \{-F_r \dots F_r\}$ and $F_r = F_g - 1$.
The laser coupling is performed for the control and target in a resonant and off-resonant fashion, respectively and results in a controlled Pauli $\widetilde{x}$ rotation of the target atom, see main-text figure 4.

The selective resonant excitation of the control atom can, for instance, be performed by first adiabatically changing the drive parameters of $H_\mathrm{DS}$ for the control atom, such that the two DS can be identified with the two SCS $|\widetilde{0}\rangle_{\mathrm{C}} = |F_g,\,F_g\rangle_{\mathrm{C}}$ and $|\widetilde{1}\rangle_{\mathrm{C}} = |F_g,\,-F_g\rangle_{\mathrm{C}}$ from which a resonant excitation can be performed, see main-text figure 4(a) upper panel. 
The adiabatic change of driving parameters is described by Eq.\eqref{eq:Hamiltonian_gate} and \eqref{eq:driveParametrization} and can, for instance, be chosen as following 
\begin{align}
\alpha(t) &= 0\notag\\
\beta(t) &= \pi/2 (1-t/T).
\end{align}
This choice ensures the two SCSs as the instantaneous DS are always antipodal with each other on the generalized Bloch sphere, which keeps the maximum separation and thus suppresses bit-flip errors and makes the adiabatic process biased, see Sec.\,\ref{E}.
Note that such an encoding is sensitive to dephasing from, \emph{e.g.} magnetic field fluctuations, but nevertheless robust against bit-flip error, thus, bias preserving.
We consider the logical state $|\widetilde{1}\rangle_{\mathrm{C}}$ to be resonantly excited to the Rydberg state $|F_r,\,-F_r\rangle_{\mathrm{C}}$, therefore, we henceforth call $\ket{\widetilde{0}}_{\mathrm{C}}$ and $|F_r,\,-F_r\rangle_{\mathrm{C}}$ the logical states of the control atom. 
Note, rotating the control qubit frame before exciting to a Rydberg state is not mandatory. One could instead directly implement a laser pulse of fixed circular polarization under a different axis of quantization, which can also perform the desired excitation.

Both the control and target Rydberg states can undergo decay due to spontaneous emission at rate $\gamma_{r}$. 
Decay of the target atom is incorporated into the simulations presented in the main-text (under the assumption that the atom decays back to the qubit manifold).
Decay of the control atom is neglected in this simplified benchmark, although we provide a more careful treatment which includes it in the next section of the appendix.

The gate-operation simulations presented in the main-text underlie the following Lindblad master equation,
\begin{equation}
\label{eq_GateFull}
\begin{aligned}
\hbar\frac{d}{dt}\hat{\rho}(t) &= -i[\hat{H}_{\mathrm{Laser}} + \hat{H}_{\mathrm{Rydberg}},\,\hat{\rho}(t)] +  \hbar\gamma_{r}\!\!\!\sum_{q=0,\pm1}\mathcal{D}\left[\sum_{m=-F_r}^{F_r}\mathcal{C}_{F_g,m;1,q}^{F_r,m+q}\ket{F_g,m-q}_{\mathrm{T}}\!\bra{F_r,m}\right]\rho,\\
\end{aligned}
\end{equation}
The coherent Hamiltonian $\hat{H}_{\mathrm{Laser}}$ contains the laser driving terms addressing the target qubit,
\begin{align}
\hat{H}_\mathrm{Laser}/\hbar= \frac{1}{2}\sum_{q = 0,\pm 1}\!\! \left(e^{ i q \Delta_r t}\, \Omega^q_r\,\hat{\mathcal{C}}^r_q + \mathrm{h.c.}\right),
\end{align}
with \mbox{$\hat{C}_q^r = \sum_m \mathcal{C}_{F_g,m; 1,q}^{F_r, m+q}|F_r,m\!+\!q\rangle_C\! \langle F_g, m |$}, equal laser Rabi frequencies $\Omega_r^{+1}=\Omega_r^{-1} = \Omega_r$, and equal-and-opposite Rydberg laser detunings $\Delta_r$.
The second part of the Hamiltonian is the Rydberg-Rydberg interaction,
\begin{equation}
\hat{H}_{\mathrm{Rydberg}}/\hbar = V \ket{F_r,\,-F_r}_{\mathrm{C}}\!\bra{F_r,\,-F_r} \otimes \!\!\!\sum_{m=-F_r}^{F_r}\!\!\!\ket{F_r,\,m}_{\mathrm{T}}\!\bra{F_r,\,m},
\end{equation}
which we assume to be a diagonal uniform density shift with strength $V$. For simplicity, we solve just this master equation without factoring in field noise via colored noise simulations.
Note for the simulations presented in the main-text we use $\gamma_r/\Omega_r = 2\pi/120$, which corresponds to a Rydberg lifetime of $1/\gamma_r = 40\,\mu s$ for $\Omega_r = 2\pi\times 3\,\mathrm{MHz}$.

The main-text Fig. 4 shows the time-evolution of different initial states under this master equation from time $t= 0 $ to $T$ with $T$ the gate time. After all operations, we assume that dissipative stabilization is applied by projecting the qubit manifold into the logical basis following Section~\ref{sec_AutonomousStabilization}. The predicted gate time, as in the main-text, is $T = \pi / \mu$ with the effective rotation strength strength of,
\begin{equation}
\mu = \frac{|\Omega_r|^2}{4 \Delta_r}\frac{1}{F_g}\frac{2F_g-1}{2F_g+1}.
\end{equation}
Note however that the optimal fidelity may be at a time $T_{\mathrm{opt}}$ slightly different from $\pi / \mu$ due to e.g. effects from finite detuning $\Delta_r$. This shift can be accounted via experimental calibration. 
Furthermore, instead of quenching on a constant laser coupling strength $\Omega_r$, we ramp it on with the following profile
\begin{align}
\Omega_r(t) = \Omega_r \left[\frac{1}{2} +\frac{ \tanh(a\cos(\pi (2t/T-1)^n))}{2\tanh(a)}\right],
\end{align}
where $a = 5$, $n = 4$, which helps to suppress left-over Rydberg population of the control atom.

We note that in general, while finite detuning from the target Rydberg state $\Delta_r$ causes error due to left-over Rydberg population, this detuning can be relatively small (e.g. $\Delta_r/\Omega_r \sim 10$) while still permitting fidelities on the order of $\sim 10^{-3}$. While the raw infidelity associated with finite $\Delta_r$ scales as $\sim \Omega_r^2/\Delta_r^2$, since this is a coherent error, it can be significantly mitigated by tailoring the gate times and/or laser ramp profiles such as the one above. In contrast, error caused by finite Rydberg interaction strength $V/\Delta_r$ (creating undesired rotation of the target) will scale as $\sim \Delta_r^2/V^2$, and cannot be as easily mitigated. For this reason we ideally we require Rydberg interaction strengths on the order of $V/\Delta_r \sim 100$ to e.g. reach the same $\sim 10^{-3}$ error thresholds that one can attain with finite detuning $\Delta_r/\Omega_r \sim 10$.

\subsection{Control atom decay correction}

\begin{figure}
\centering
\includegraphics[width=1\textwidth]{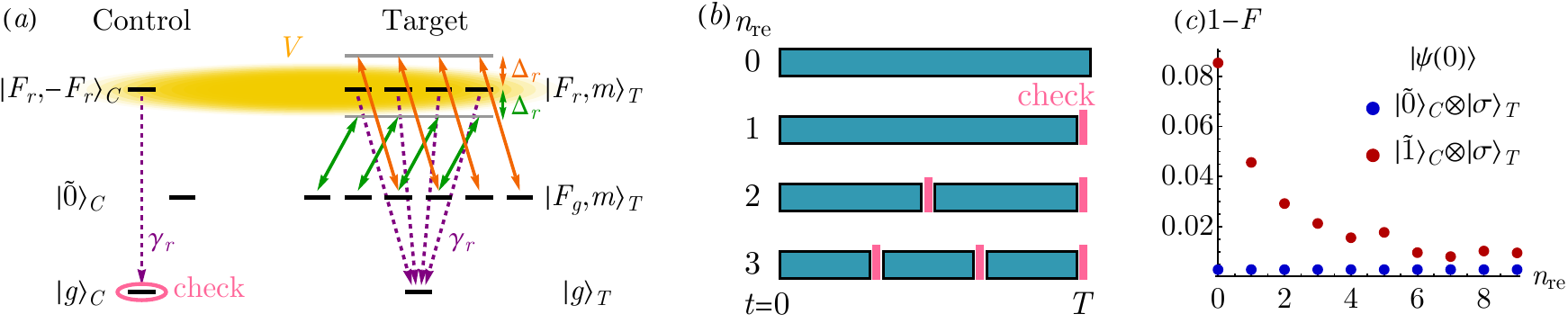}
\caption{(a) Schematic of the Hilbert space used for numerical simulation of the bias-preserving $\hat{\mathrm{C}}_X$ gate with more careful accounting of Rydberg decay due to spontaneous emission. (b) Schematic of the monitoring protocol that makes $n_{\mathrm{re}}$ checks of the $\ket{g}_{\mathrm{C}}$ population during the gate to detect control Rydberg decay. (c) Bit-flip infidelity of the gate, depending on whether the control atom starts in $\ket{\widetilde{0}}_{\mathrm{C}}$ or $\ket{\widetilde{1}}_{\mathrm{C}}$ (the latter corresponds to $\ket{F_r,-F_r}_{\mathrm{C}}$). We vary the number $n_{\mathrm{re}}$ of checks for control atom decay. The parameters used are $F_g = 4$, $\Delta_r / \Omega_r = 2$, $\gamma_r / \Omega_r = 2\pi/120$, and $V/\Omega_r = 100$. Note that during each interval we use a constant Rabi frequency $\Omega_r$ (rather than the ramp in the previous section) for simplicity.}
\label{fig:CXGateSupplement}
\end{figure}

In this section we introduce a protocol for the $\hat{\mathrm{C}}_X$ gate to correct non-bias preserving processes due to control atom decay from the Ry state, which extends the discussion of the prior section. 
In particular, here we include decay of the control atom, assume that both atoms' Rydberg states decay into an independent ground state rather than the qubit manifold and provide a means of correcting control atom decay by monitoring the ground state population.

The Hilbert space we consider for the protocol is shown in Fig~\ref{fig:CXGateSupplement}(a). 
The target atom still consists of a ground and Rydberg manifold $\ket{F_g,m}_{\mathrm{T}}$ and $\ket{F_r,m}_{\mathrm{T}}$, respectively, but now also an additional ground state $\ket{g}_{\mathrm{T}}$ into which the Rydberg state decays. 
The (hyper-)fine structure of this additional ground state is not tracked. 
Moreover, the control atom likewise has qubit and Rydberg states $\ket{\widetilde{0}}_{\mathrm{C}}$ and $\ket{F_r,-F_r}_{\mathrm{C}}$, as well as a ground state $\ket{g}_{\mathrm{C}}$. 
The laser coupling Hamiltonian $\hat{H}_{\mathrm{Laser}}$ and the Rydberg-Rydberg interaction $\hat{H}_{\mathrm{rydberg}}$ remain the same, but the master equation is adjusted to be
\begin{equation}
\label{eq_GateFullAdjust}
\begin{aligned}
\hbar\frac{d}{dt}\rho &= -i[\hat{H}_{\mathrm{laser}} + \hat{H}_{\mathrm{rydberg}},\rho] + \hbar\gamma_{r}\mathcal{D}[\ket{g}_{\mathrm{C}}\bra{F_r,-F_r}]\rho + \hbar\gamma_{r}\sum_{m=-F_r}^{F_r} \mathcal{D}[\ket{g}_{\mathrm{T}}\bra{F_r,m}]\rho.
\end{aligned}
\end{equation}

With control atom decay incorporated, we observe that if such a decay happens during gate execution, Rydberg blockade will cease and the target will start to rotate, which can cause non-negligible bit-flip error. 
To mitigate this effect, we can adjust our gate protocol to include a monitoring process. 
This adjusted protocol periodically checks whether the control atom decayed during the gate time by measuring the ground state.

The protocol proceeds by splitting the full gate time $T$ into $n_{\mathrm{re}}$ equal subintervals of length $T/n_{\mathrm{re}}$ indexed by $n\in \{1,2, \dots n_{\mathrm{re}}\}$, as depicted in Fig.~\ref{fig:CXGateSupplement}(b). At the end of each subinterval, we check if Rydberg decay of the control has occurred by projectively measuring the population of $\ket{g}_{\mathrm{C}}$. The case $n_{\mathrm{re}}=0$ means we don't perform any checks. If the control atom is detected in the ground state after a given check, it should have been in the Rydberg state and thus blockading rotation of the target. Upon such detection we stop the gate operation early (turn off the lasers addressing the target) to prevent further unwanted rotation. If no decay was detected, the gate continues to the next subinterval. At the end of the gate we assume that any direct population outside the target's qubit manifold is lost, then finally we apply dissipative stabilization.

For simplicity, to benchmark this augmented protocol we simply compute the gate fidelity for certain initial states. The initial condition we take for the gate evolution is a pure state,
\begin{equation}
\ket{\psi(0)} = \ket{\sigma'}_{\mathrm{C}}\otimes \ket{\sigma}_{\mathrm{T}},\\
\end{equation}
with $\sigma, \sigma' \in \{\widetilde{0}, \widetilde{1}\}$, noting that here $\ket{\widetilde{1}}_{\mathrm{C}} = \ket{F_r,-F_r}_{\mathrm{C}}$.

The ideal final state after the gate operation $\ket{\psi_{\mathrm{ideal}}}$ depends on whether the control was Rydberg blockading or not:
\begin{equation}
\ket{\psi_{\mathrm{ideal}}} = \begin{cases}
\ket{\widetilde{1}}_{\mathrm{C}} \otimes \ket{\sigma}_{\mathrm{T}}, & \ket{\psi(0)} = \ket{\widetilde{1}}_{\mathrm{C}}\otimes \ket{\sigma}_{\mathrm{T}}\\
\ket{\widetilde{0}}_{\mathrm{C}} \otimes \ket{\overline{\sigma}}_{\mathrm{T}},  & \ket{\psi(0)} = \ket{\widetilde{0}}_{\mathrm{C}}\otimes \ket{\sigma}_{\mathrm{T}}\\
\end{cases}
\end{equation}
where $\overline{\widetilde{0}} = \widetilde{1}$, $\overline{\widetilde{1}} = \widetilde{0}$. The fidelity is obtained via,
\begin{equation}
\label{eq_FidelityUhlmannCX}
F =  \text{tr} \left[\rho_f\cdot\ket{\psi_{\mathrm{ideal}}}\bra{\psi_{\mathrm{ideal}}}\right].
\end{equation}

In Fig~\ref{fig:CXGateSupplement}(c) we plot the resulting fidelity for different numbers of checks $n_{\mathrm{re}}$. If the control starts in $\ket{\widetilde{0}}_{\mathrm{C}}$, there is no Rydberg blockade or decay. The fidelity only depends on $\Omega_r/\Delta_r$, approaching $F = 1$  in the limit $\Omega_r/\Delta_r \to 0$ (for which adiabatic elimination of the target Rydberg state is perfect). On the other hand, if the control starts in $\ket{r}_{\mathrm{C}}$, the fidelity starts out poor due to Rydberg decay of the control, but improves exponentially in the number of checks $n_{\mathrm{re}}$, approaching the fidelity of the non-blockaded case in the limit $n_{\mathrm{re}}\to \infty$. Note, the non-monotonic behavior occurs due to finite detuning $\Delta_r$ causing target Rydberg state population oscillations. Crucially, these simulations show that with only a few checks, the effective bit-flip fidelity can be greatly decreased.

This augmented procedure does not correct for phase-flip error, which will generally occur at rate $\sim 1 - \exp(-\gamma_r T) \sim \gamma_r T$ (for $\gamma_r T \ll 1$). However, this type of error can still be rendered small with longer-lived Rydberg states and shorter gate times. Reduction of the bit-flip fidelity compared to the phase-flip also remains favorable for biased error correction protocols.

\section{Non-bias-preserving gates: Universality on the physical qubit level}

As a supplement, in this section we provide several gate designs for the dark spin-cat encoding, which could be more native but lose the bit-flip protection. We first explain the collision-and-expansion trick and how to use this for the $\hat{U}_{\widetilde{x}}(\alpha_x)$ gate design, which is inspired by the bosonic-cat code gate construction as proposed in Ref.~\cite{albert2016holonomic}. This gate does not preserve the error bias as we have to reduce the spatial separation for the two SCSs on the generalized Bloch sphere in order to exchange the population between $\ket{\widetilde{0}}$ and $\ket{\widetilde{1}}$ code states. 
This gate, together with 1-qubit $\hat{U}_{\widetilde{z}}(\alpha_z)$ and entangling $\hat{\mathrm{C}}_Z$ operation, which we discussed before, complete a universal gate set on the physical encoding level. 
Additionally, in the second part of the section, we discuss a simplified unbiased construction of the $\hat{\mathrm{C}}_X$ gate using the collision-and-expansion trick.

Here are two things that we would like to point out. First, in principle, those unbiased gates are not strictly necessary, as the bias-preserving gate set shown in the main text (including $\hat{\mathrm{C}}_X$ and Toffoli) are universal for fault-tolerant operations logical (repetition code) qubits~\cite{guillaud2019repetition}. 
Nevertheless, those non-bias-preserving constructions can still offer sufficient advantages when the performance of bias-preserving $\hat{\mathrm{C}}_X$ gates is limited. 
For example, neutral atom quantum processors exhibit negligible 1-qubit gate infidelities as compared to the $\hat{\mathrm{C}}_Z$ gate performance, which is limited by Rydberg decay. 
However, Ref.~\cite{sahay2023high} pointed out that, if it is possible to detect the location of the decay event and supply a fresh atom initialized in $\ket{1}$ (since only $\ket{1}$ is excited to the  Rydberg state), it is convertable to a dephasing error with known location.
This so-called ``biased-erasure'' error during $\hat{\mathrm{C}}_Z$ implementation provides an error correction threshold around 10\% without the need of bias-preserving $\hat{\mathrm{C}}_X$ gate, and can be even higher exploiting ideas presented in~\cite{sahay2023high}. 
As explained in the main text, our dark spin-cat encoding can be well adapted to this framework.

\subsection{Holonomic 1-qubit $\hat{U}_{\widetilde{x}}(\alpha_x)$ gate}

\begin{figure}
    \centering
    \includegraphics[width=0.5\linewidth]{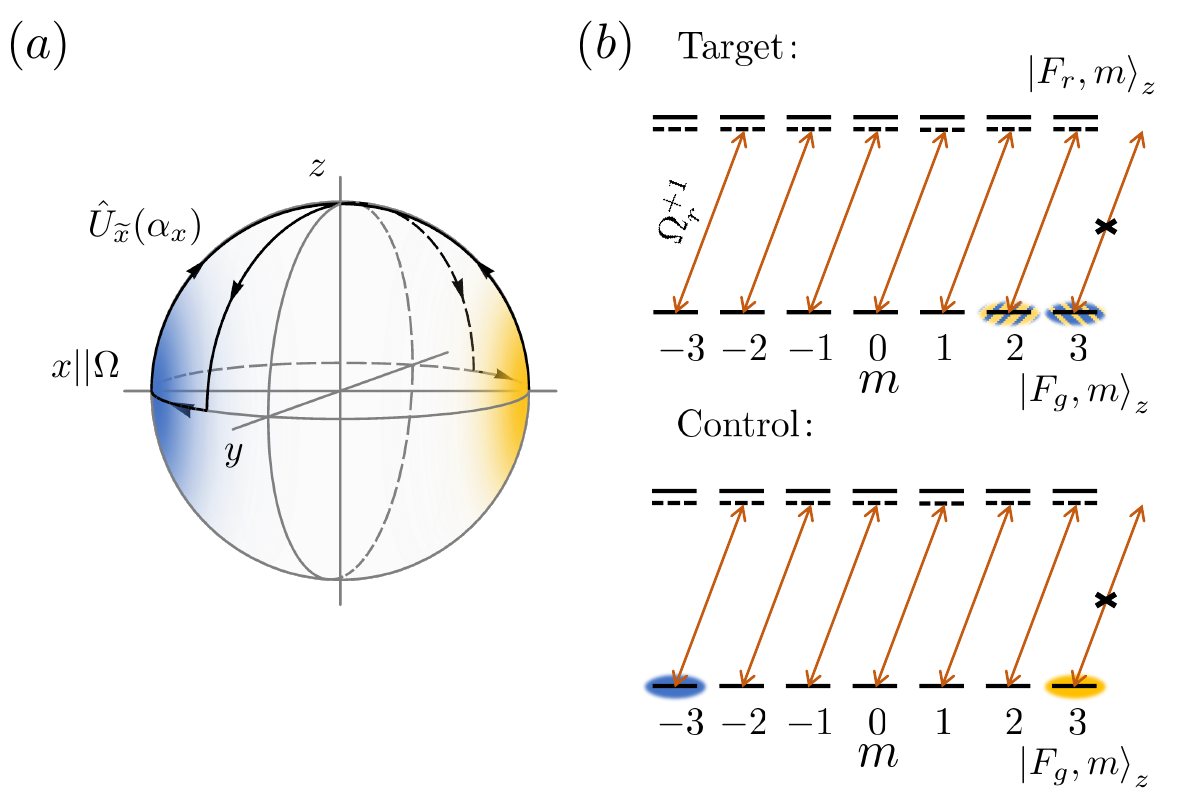}
    \caption{Schematic plot for native non-bias-preserving gate design. (a) The path of two SCSs during $\hat{U}_{\widetilde{x}}(\alpha_x)$ gate execution. They first collide and then separate in another direction. (b) The entangling operation in the middle of the native $\hat{\mathrm{C}}_X$ execution. For the target atom, the collision step will convert $\ket{\widetilde{+}}$ to $\ket{F_g, F_g}_z$ and $\ket{\widetilde{-}}$ to $\ket{F_g, F_g-1}_z$.}
    \label{fig:SM_Uxy}
\end{figure}

In this section, we briefly present how to construct a single-qubit $\hat{U}_{\widetilde{x}}(\alpha_x)$ gate by adiabatically changing the driving parameters $\Omega_{\pm 1}(t)$. 
We start with $\Omega_{+1} = -\Omega_{-1} = \Omega$, such that the code subspace is spanned by $\{\ket{F_g, F_g}_x, \ket{F_g, -F_g}_x\}$, \emph{i.e.} two SCSs pointing along the $x$-axis on the generalized Bloch sphere. 
We first adiabatically reduce the amplitude $\Omega_{-1}$ to $0$ in order to collide the two SCSs at the north pole of the generalized Bloch sphere. 
Then, we separate them in another direction by adiabatically increasing $\Omega_{-1}$ to $-\Omega e^{-2i\alpha_x}$, and finally changing $\Omega_{-1}$ to $-\Omega$ without affecting its amplitude. 
The location change for the two SCSs during the whole process is illustrated in Fig.\,\ref{fig:SM_Uxy}.

Note, in the absence of the $\Omega_0$ drive, one of the dark states lies in the subspace spanned by $\{\ket{F_g, F_g-2k}_z\}$ while another one is in the subspace spanned by $\{\ket{F_g, F_g-(2k+1)}_z\}$. 
Therefore, during the first step where we adiabatically turn off $\Omega_{-1}$, the state $\frac{1}{\sqrt{2}}(\ket{F_g, F_g}_x + e^{2i\pi F_g}\ket{F_g, -F_g}_x)$ connects to $\ket{F_g, F_g}_z$, while $\frac{1}{\sqrt{2}}(\ket{F_g, F_g}_x - e^{2i\pi F_g}\ket{F_g, -F_g}_x)$ is connected to $\ket{F_g, F_g-1}_z$. 
Following the convention in Ref.~\cite{albert2016holonomic}, we denote this mapping as $S_0^\dagger$, Due to the rotational symmetry in the system the operation for the second step is  $S_{\alpha_x}$, which is related to $S_0$ by a rotation operator $S_{\alpha_x} = \mathcal{R}(\alpha_x,0,0) S_0  \mathcal{R}^\dagger(\alpha_x,0,0)$. The third step is simply a rotation $\mathcal{R}^\dagger(\alpha_x,0,0)$. The whole process leads to an operation $\mathcal{R}^\dagger(\alpha_x,0,0) S_{\alpha_x} S_0^\dagger = S_0 \mathcal{R}^\dagger(\alpha_x,0,0) S^\dagger_0$ on the dark-state subspace. Therefore, $\frac{1}{\sqrt{2}}(\ket{F_g, F_g}_x + e^{2i\pi F_g}\ket{F_g, -F_g}_x)$ will pick up a $e^{i\alpha_x F_g}$ phase while $\frac{1}{\sqrt{2}}(\ket{F_g, F_g}_x + e^{2i\pi F_g}\ket{F_g, -F_g}_x)$ will get a phase of $e^{i\alpha_x (F_g-1)}$. As a result, an arbitrary superposition of the two code states will become the following state after the whole process
\begin{equation}
    c_+ \frac{\ket{\widetilde{0}} + e^{2i\pi F_g}\ket{\widetilde{1}}}{\sqrt{2}} + c_- \frac{\ket{\widetilde{0}} - e^{2i\pi F_g}\ket{\widetilde{1}}}{\sqrt{2}} \longrightarrow e^{i\alpha_x F_g} \left[c_+ \frac{\ket{\widetilde{0}} + e^{2i\pi F_g}\ket{\widetilde{1}}}{\sqrt{2}} + c_- e^{-i\alpha_x}\frac{\ket{\widetilde{0}} - e^{2i\pi F_g}\ket{\widetilde{1}}}{\sqrt{2}} \right],
\end{equation}
which corresponds to $\hat{U}_{\widetilde{x}}(\pm\alpha_x)$ for a half-integer / integer $F_g$. This operation, together with $\hat{U}_{\widetilde{z}}(\alpha_z)$ gate, provides arbitrary 1-qubit rotation on the dark spin-cat encoding.

\subsection{Native implementation for non-bias-preserving $\hat{\mathrm{C}}_X$ gate}

With arbitrary 1-qubit rotations and the $\hat{\mathrm{C}}_Z$ entangling gate, we can achieve universal operations on the physical encoding level, including the $\hat{\mathrm{C}}_X$ operations as $\hat{\mathrm{C}}_X = \hat U_{\widetilde{y},T}(\frac{\pi}{2}) \hat{\mathrm{C}}_Z \hat U_{\widetilde{y},T}(-\frac{\pi}{2})$. However, we can further simplify the $\hat{\mathrm{C}}_X$ control sequence in a more native manner using the collision-and-expansion trick we introduced in the $\hat{U}_{\widetilde{x}}(\alpha_x)$ gate construction.
 
We briefly explain the idea as follows. The first step is again converting the encoding from Fock states in $x$ basis to Fock in $z$ basis, but for the control and target atom the protocol is different. For the control atom, the conversion protocol can be the same as that used for the control atom in bias-preserving $\hat{\mathrm{C}}_X$ or $\hat{\mathrm{C}}_Z$ construction shown in the main text, i.e., adiabatically transfer $\ket{\widetilde{0}/\widetilde{1}} = \ket{F_g, \pm F_g}_x$ to $\ket{F_g, \pm F_g}_z$ while always keeping them antipodal during the evolution. On the other hand, for the target atom we can use the collision trick that converts $\ket{\widetilde{+}}$ to $\ket{F_g, F_g}_z$ and $\ket{\widetilde{-}}$ to $\ket{F_g, F_g-1}_z$ by adiabatically turning off the $\Omega_{-1}$ drive \footnote{This convention applies when $F_g$ is an integer. If $F_g$ is a half-integer, then the protocol will map $\ket{\widetilde{-}}$ to $\ket{F_g, F_g}_z$ and $\ket{\widetilde{+}}$ to $\ket{F_g, F_g-1}_z$.}. 
 
After the conversion above, we need to perform the entangling operation. Here we consider a Rydberg manifold with $F_r = F_g$. Consider using a $\sigma_+$ polarized drive to address the state from the encoded manifold to the Rydberg manifold. Due to the selectivity that comes from the polarization, $\ket{F_g, F_g}_z$ will not couple with the Rydberg levels, but $\ket{F_g, -F_g}_z$ and $\ket{F_g, F_g-1}_z$ will. Therefore, one can use the standard $\hat{\mathrm{C}}_Z$ control sequence~\cite{levine2019parallel} so that both $\ket{F_g, F_g}_{z,C} \otimes \ket{F_g, F_g-1}_{z,T}$ and $\ket{F_g, -F_g}_{z,C} \otimes \ket{F_g, F_g}_{z,T}$ will pick up a phase $e^{i\phi}$ while $\ket{F_g, -F_g}_{z,C} \otimes \ket{F_g, F_g-1}_{z,T}$ will get a phase $e^{i(2\phi-\pi)}$. This is equivalent to a $\hat{\mathrm{C}}_Z$ operation up to single-qubit phase rotations. Finally, we need to map the encoded subspaces from the Fock $z$ basis back to the Fock $x$ basis. We can choose a different trajectory compared with that used in the first step to account for the required extra 1-qubit gates. For the control atom, we still need to keep the two SCSs antipodal, which is similar to the idea used in $\hat{U}_{\widetilde{z}}(\alpha_z)$ gate construction. On the other hand, for the target atom, we can follow the 2nd and 3rd steps used in $\hat{U}_{\widetilde{x}}(\alpha_x)$ gate design. At the end of the protocol, only $\ket{\widetilde{1}}_C \otimes \ket{\widetilde{-}}_T$ will get a relative $\pi$ phase, which results in a $\hat{\mathrm{C}}_X$ gate. 

Finally, we would like to point out that this simplified version of $\hat{\mathrm{C}}_X$ execution can still be adapted to the ``biased-erasure'' framework, provided that the error from 1-qubit control is negligible and the Rydberg decay during entangling operations can be detected. If the decay is detected on the control atom, we prepare a fresh $\ket{\widetilde{1}}$, and if it is on the target atom we prepare a fresh $\ket{\widetilde{-}}$. In this way, we again have the knowledge about both the location and the type of the error (for the control atom the error is $Z$ type, while for the target it is $X$ type). This noise pattern is equivalent to the $\hat{\mathrm{C}}_X = \hat U_{\widetilde{y},T}(\frac{\pi}{2}) \hat{\mathrm{C}}_Z \hat U_{\widetilde{y},T}(-\frac{\pi}{2})$ construction proposed in~\cite{sahay2023high} where the error during $\hat{\mathrm{C}}_Z$ in the middle is biased-erasure that both the control and target will have $Z$ type of error after erasure conversion. $Z$ error on the target will be converted to $X$ error after $\hat U_{\widetilde{y},T}(\frac{\pi}{2})$ rotation, which leads to the same error structure for the $\hat{\mathrm{C}}_X$ construction as we proposed here.

\twocolumngrid

\bibliography{bib}
\end{document}